%% file: main.tex
\documentclass[numbers]{article}

\usepackage[preprint]{neurips_2024}

\usepackage[utf8]{inputenc} 
\usepackage[T1]{fontenc}    
\usepackage{hyperref}       
\usepackage{url}            
\usepackage{booktabs}       
\usepackage{amsfonts}       
\usepackage{nicefrac}       
\usepackage{microtype}      
\usepackage{xcolor}         

\usepackage{graphicx}
\usepackage[table]{xcolor}
\usepackage{todonotes}
\usepackage{menukeys}
\usepackage{natbib}
\usepackage{gensymb}
\usepackage{amsmath}
\usepackage{makecell}
\usepackage{booktabs}
\usepackage{tabularx}
\usepackage{multirow}

\title{AI-Guided Discovery of Novel Ionic Liquid Solvents for Industrial CO${_2}$ Capture }

\author{%
Davide Garbelotto\textsuperscript{1}, 
Alexander Lobo\textsuperscript{1}, 
Urvi Awasthi\textsuperscript{1}, 
Oleg Medvedev\textsuperscript{1}, 
Srayanta Mukherjee\textsuperscript{1}, 
Anton Aristov\textsuperscript{1}, 
Konstantin Polunin\textsuperscript{1}, 
Alex De Mur\textsuperscript{1}, 
Leonid Zhukov\textsuperscript{1}, 
Azad Huseynov\textsuperscript{2,3}, 
Murad Abdullayev\textsuperscript{2,3} 
\\[2ex]
\textsuperscript{1}BCG X AI Science Institute \\
\textsuperscript{2}Caspian AI Institute for Energy Transition \\
\textsuperscript{3}State Oil Company of the Republic of Azerbaijan (SOCAR)
}

\begin{document}

\maketitle


\input{abstract}

\input{sections/00_intro}
\input{sections/01_background}
\input{sections/02_methods}
\input{sections/03_results}

\input{sections/99_conclusions}

\input{sections/ack} 
\bibliography{bibliography}


\appendix
\input{sections/999_appendix}


\newpage

\end{document}

%% file: abstract.tex
\begin{abstract}
We present an AI-driven approach to discover compounds with optimal properties for CO${_2}$ capture from flue gas—refinery emissions’ primary source. Focusing on ionic liquids (ILs) as alternatives to traditional amine-based solvents, we successfully identify new IL candidates with high working capacity, manageable viscosity, favorable regeneration energy, and viable synthetic routes. Our approach follows a five-stage pipeline. First, we generate IL candidates by pairing available cation and anion molecules, then predict temperature- and pressure-dependent CO${_2}$ solubility and viscosity using a GNN-based molecular property prediction model. Next, we convert solubility to working capacity and regeneration energy via Van’t Hoff modeling, and then find the best set of candidates using Pareto optimization, before finally filtering those based on feasible synthesis routes. We identify 36 feasible candidates that could enable 5–10\% OPEX savings and up to 10\% CAPEX reductions through lower regeneration energy requirements and reduced corrosivity—offering a novel carbon-capture strategy for refineries moving forward.

\end{abstract}

%% file: sections/00_intro.tex
\begin{figure}[!h]
    \centering
    \includegraphics[width=1\linewidth]{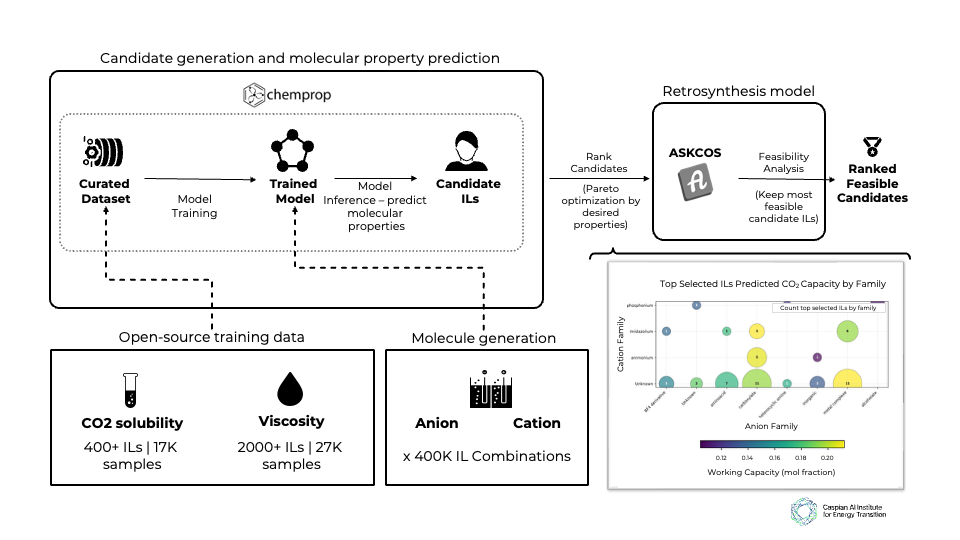}
    \caption{Overview of AI-driven IL candidate screening framework}
    \label{fig:overviewabstract}
\end{figure}

\section{Introduction}
\label{sec:intro}

Refinery operations are critical to global energy systems, yet they come with substantial environmental implications, particularly the significant emissions of carbon dioxide (CO$_2$) \cite{Boot-Handford2014, Rochelle2009}. As global efforts intensify to combat climate change, the challenge to economically and sustainably reduce these emissions becomes increasingly urgent. Traditional carbon capture methods, heavily reliant on conventional solvents like monoethanolamine (MEA) and methyldiethanolamine (MDEA), have served as workhorses for decades \cite{Boot-Handford2014, Rochelle2009, Duby2022}. However, they come with notable limitations—high energy demands for regeneration, significant solvent losses, corrosivity, and substantial operating and capital expenditures (OPEX and CAPEX) \cite{Rochelle2009}. These constraints not only diminish economic efficiency but also present critical environmental concerns, emphasizing the urgent need for innovation in carbon capture technologies.

In response to these challenges, ionic liquids (ILs) emerge as a groundbreaking class of solvents, distinguished by their negligible volatility, high thermal stability, remarkable CO$_2$ selectivity, and unparalleled structural tunability \cite{Brennecke2010,Wappel2010,Dai2017}. Unlike traditional solvents, ILs offer an extensive chemical space that can be strategically explored to optimize their physicochemical properties for specific industrial scenarios \cite{Dai2017}. This vast combinatorial potential positions ILs to revolutionize CO$_2$ capture processes, delivering transformative reductions in energy consumption, equipment corrosion, solvent loss, and ultimately operational costs.

Despite their extraordinary promise, the systematic exploration of ILs to identify candidates optimized for refinery CO$_2$ capture is an immense task, impossible to navigate efficiently through traditional experimental methods alone. This is precisely where artificial intelligence (AI) steps in. Recent advances in AI-driven property prediction and material discovery frameworks have unlocked capabilities to rapidly and accurately screen large molecular spaces \cite{Yang2019,Gilmer2017,Coley2019}. By integrating predictive modeling techniques, such as graph neural networks, with systematic chemical exploration and thermodynamic inference, AI can swiftly pinpoint ionic liquids with superior performance profiles \cite{Yang2019}. These computational approaches dramatically accelerate solvent discovery, transforming what was once an intractable task into a precise, data-driven exploration.

In this work, we leverage cutting-edge AI methods to rigorously and efficiently identify ionic liquids tailored explicitly for CO$_2$ capture in refinery operations. 
Our approach addresses the pressing dual objectives of reducing emissions and enhancing operational efficiency, opening the door to substantial environmental and economic advancements within the refining industry.

%% file: sections/01_background.tex
\section{Background and Related Work}

Chemical absorption methods using solvents such as monoethanolamine (MEA) and methyldiethanolamine (MDEA) have been the standard approach in industrial CO$_2$ capture for decades \cite{Boot-Handford2014,Rochelle2009}. These conventional solvents have demonstrated effectiveness in selectively absorbing CO$_2$ from industrial flue gases. However, significant limitations persist, notably high energy requirements for solvent regeneration, corrosion issues due to solvent degradation, and considerable solvent losses due to high volatility \cite{Chi2002,Davis2009}. Such factors significantly contribute to elevated operational (OPEX) and capital expenditures (CAPEX), thereby motivating the pursuit of alternative solvents with improved economic and environmental profiles.

Ionic liquids (ILs), salts existing in a liquid state below 100\degree C, represent a promising alternative solvent class, overcoming several limitations associated with conventional solvents \cite{Brennecke2010,Dai2017}. These ILs exhibit negligible vapor pressures, superior thermal stability, and exceptional tunability through systematic alterations in their constituent ions \cite{Dai2017}. Research into ILs has highlighted their capacity to reduce energy demands for regeneration significantly and mitigate equipment corrosion issues due to their inherent non-volatility and chemical stability \cite{Chi2002}. The large chemical space provided by varying combinations of cations and anions offers significant potential for developing tailored solutions optimized for specific industrial CO$_2$ capture applications.

Typical refinery carbon capture setups employ a two-tower absorption-desorption system. In the absorption (scrubber) tower, solvent contacts the flue gas at near-ambient conditions—typically around 40\degree C and atmospheric pressure—facilitating maximum CO$_2$ uptake \cite{Rochelle2009,Boot-Handford2014}. Subsequently, the CO$_2$-rich solvent undergoes regeneration in the desorption tower, operating at higher temperatures around 100-120\degree C and pressures around 1-2 bar. At these elevated conditions, CO$_2$ is effectively released, and the solvent is regenerated for reuse.

The performance and efficiency of carbon capture solvents under these industrial conditions are critically dependent on several key properties, including CO$_2$ working capacity, solvent viscosity, thermal stability, and the required energy for solvent regeneration. The harsh thermal cycling and operational demands of industrial absorption-desorption systems necessitate solvents with robust performance characteristics, significantly influencing the practical and economic viability of carbon capture technologies in refineries.

\begin{figure}[t]
    \centering
    \includegraphics[width=1\linewidth]{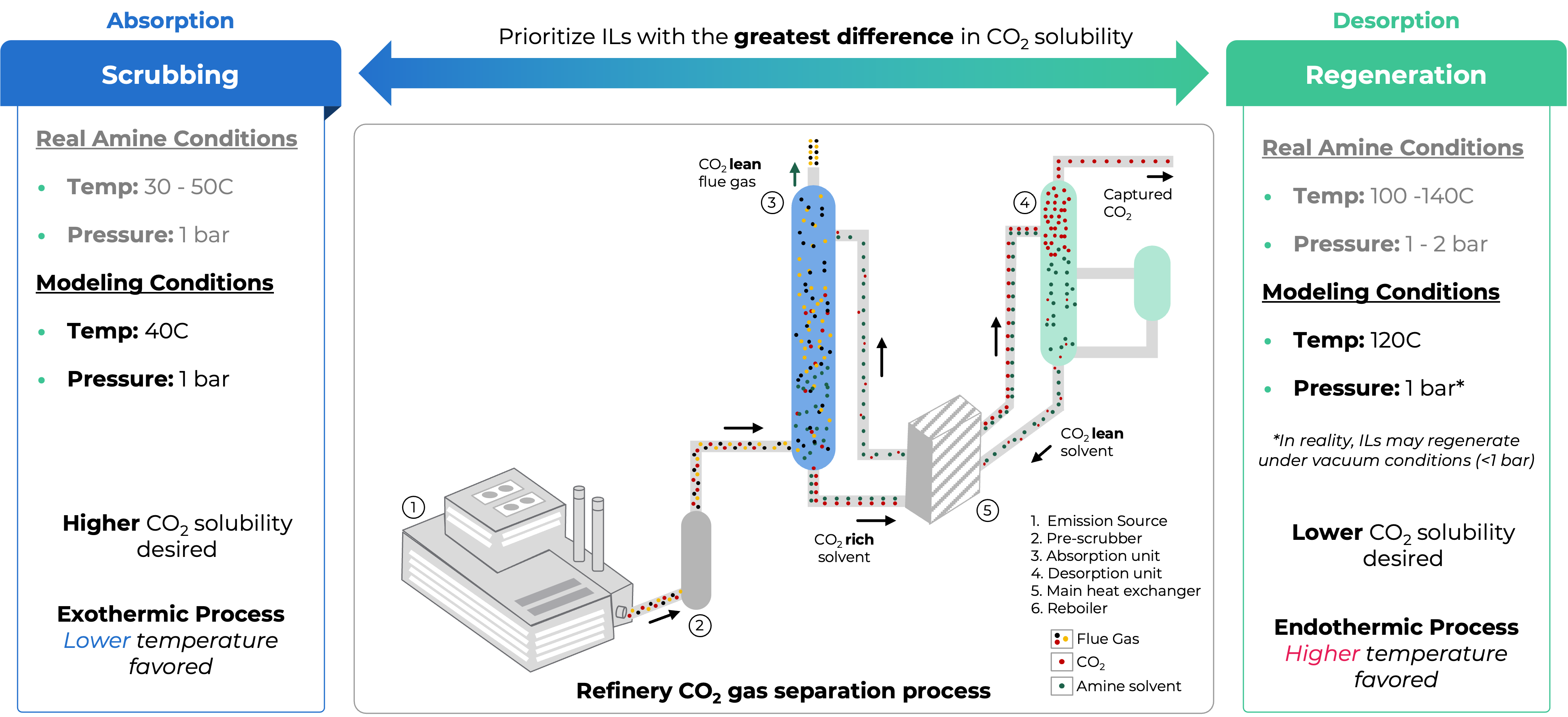}
    \caption{Schematic of absorption–desorption process for CO$_2$ capture in industrial scrubbing systems. Absorption occurs at low temperature and pressure, while regeneration requires elevated thermal conditions to release CO$_2$ and recycle solvent}
    \label{fig:operating-conditions}
\end{figure}

In response to the considerable experimental challenges presented by the large chemical landscape of ILs, artificial intelligence (AI)-based methods have emerged as transformative tools in material discovery \cite{Yang2019,Gilmer2017,Coley2019}. Recent advancements in AI, particularly through the use of graph neural networks (GNNs), have demonstrated substantial potential in rapidly screening and predicting optimal candidates from extensive chemical spaces \cite{Yang2019}. These AI-driven frameworks efficiently predict crucial physicochemical properties, thus significantly accelerating the solvent selection and optimization processes \cite{Gilmer2017,Coley2019}.

However, despite these advances, the application of AI-driven approaches explicitly tailored to identify and evaluate ILs for refinery-specific CO$_2$ capture under realistic absorption-desorption conditions remains relatively unexplored. The current research thus aims to bridge this gap, leveraging AI methods to systematically explore the potential of ILs to meet the stringent demands of industrial carbon capture operations, ultimately advancing the sustainability and economic viability of refinery processes.

%% file: sections/02_methods.tex
\section{Methods}
\label{sec:methods}

\subsection{Framework Overview}
\label{sec:framework}
We aim to model the temperature- and pressure-dependent physical properties of ionic liquids (ILs)—specifically CO$_2$ solubility and viscosity—as a function of their molecular structure, represented via canonical SMILES strings. These models are trained on curated experimental datasets and used to predict properties at unobserved thermodynamic conditions and for novel IL combinations. The ultimate objective is to deploy this learned predictive framework as a generative engine for prioritizing IL candidates for post-combustion CO$_2$ capture in refinery operations.

While a growing body of experimental work has reported solubility and viscosity data for ILs under controlled laboratory conditions, the coverage of temperature–pressure space remains sparse, and many IL combinations remain unexplored or partially characterized \cite{Mehrkesh2016, D3DD00040K, data4020088, CHEN2025335, Dong2007}. In this context, we seek to generalize beyond observed inputs by learning structure–property relationships that allow rapid inference across untested conditions and novel IL chemistries. This is particularly relevant for deployment in real-world refinery scenarios, where practical operating conditions (e.g., 1–2 bar, 40–120$^{\circ}$C) may not align with the available measurements in the literature.
To address this, we construct a modular prediction and screening pipeline composed of five primary stages:
\begin{enumerate}
    \item   \textbf{Ionic Liquid Candidate Generation.} We generate a set of over 400,000 IL candidates by combinatorially pairing curated cations and anions from various IL property datasets.
    \item   \textbf{GNN-based Molecular Property Prediction.} We train directed message passing neural network (D-MPNN) models to predict solubility (mol CO$_2$/total mols) and viscosity (mPa·s). These models take the separated SMILES of the cation and anion, temperature and pressure as inputs to predict each target property.
    \item   \textbf{Physics-Based Thermodynamic Property Estimation.} From predicted solubility values over a range of temperatures, we derive CO$_2$ working capacity and regeneration energy for each IL via Van’t Hoff analysis. The working capacity reflects usable CO$_2$ absorption under operating (rich) and regeneration (lean) conditions, while the slope of the Van’t Hoff fit provides a proxy for the thermal energy penalty of solvent regeneration.
    \item \textbf{Pareto-Based Candidate Selection.} We apply multi-objective Pareto front analysis to identify ILs that optimally balance working capacity (maximize), CO$_2$ solubility at scrubbing conditions (maximize), and viscosity at full loading (minimize). This approach surfaces non-dominated candidates without imposing arbitrary trade-offs and supports interpretable exploration of process-relevant performance boundaries.
    \item   \textbf{Synthesis Feasibility Filtering.} Finally, we apply automated retrosynthesis tools to filter candidates with no known or low-complexity synthesis routes to ensure that top-performing ILs are not only thermodynamically promising but also synthetically accessible.
\end{enumerate}

An overview of the pipeline is illustrated in Figure \ref{fig:overviewabstract}, which shows the integration of model-based inference, thermodynamic post-processing, and feasibility-aware ranking. Importantly, this framework supports screening at scale: property predictions can be made for hundreds of thousands of IL candidates in silico, enabling expansive exploration of the chemical design space without relying on high-throughput experimental workflows.
While each stage of this process introduces uncertainty—both from data noise and model generalization error—we argue that combining learned representations with physically motivated scoring allows for robust and interpretable downstream decision-making. This view reflects recent findings in the literature that hybrid ML–thermodynamic models can outperform purely empirical or purely physics-based baselines across molecular property tasks \cite{D3DD00040K, Vermeire_2021}.

In contrast to prior work which either focuses on predicting properties outside of refinery operating conditions or imposes handcrafted rules for IL screening \cite{Racki2025}, our pipeline offers a data-driven yet extensible solution for identifying ILs that are effective, efficient, and viable for carbon capture under real-world refinery constraints.

\subsection{Ionic Liquid Candidate Generation}
\label{sec:il-gen}

We generated 405{,}891 ionic liquid (IL) candidates by combinatorially pairing cations and anions from cleaned datasets outlined in Table \ref{tables/data-summary}. While this naive Cartesian approach constrains the search space to known ions, the framework is fully modular and can incorporate any generative strategy, including rule-based design, structural editing, or generative models to expand chemical diversity.

\subsection{GNN-based Molecular Property Prediction}
\label{sec:dmpnn}

To predict temperature- and pressure-dependent CO$_2$ solubility and viscosity for a broad space of ionic liquids (ILs), we employ the Chemprop directed message passing graph neural network (D-MPNN) framework \cite{Heid2023}. 

\subsubsection{Directed message passing neural network (D-MPNN) architecture}

Our predictive models follow the directed message passing formulation implemented in Chemprop. Each ion (cation or anion) is represented as a molecular graph $G=(V,E)$ with atoms as nodes and directed bonds as edges. For each directed edge $e_{ij}$, we initialize a hidden state $h_{ij}^{(0)}$ from standard atom/bond features (e.g., bond type, aromaticity, hybridization).

During message passing, directed edge states are updated for $T$ steps via neighbor aggregation that avoids immediate backtracking:
\[
h_{ij}^{(t+1)} = \mathrm{ReLU}\!\left( W_m \sum_{k \in \mathcal{N}(i)\setminus j} h_{ki}^{(t)} + b_m \right).
\]

After message passing, edge states are aggregated into atom representations and pooled to obtain a graph-level embedding:
\[
h_{\mathrm{mol}} = \mathrm{Aggregate}\left(\{h_i\}_{i\in V}\right),
\]
where $\mathrm{Aggregate}(\cdot)$ is typically a sum or mean.

We compute separate embeddings for the cation and anion and concatenate them to form the ionic-liquid representation:
\[
h_{\mathrm{IL}} = [h_{\mathrm{cation}} \,\|\, h_{\mathrm{anion}}].
\]
This vector is concatenated with the thermodynamic condition features (temperature and pressure) and passed to an MLP regression head to predict each target property.



We train two separate single-property prediction tasks to predict CO$_2$ solubility (in mol CO$_2$/total mols) and dynamic viscosity (in mPa·s) using the same D-MPNN architecture for each. We include temperature and pressure conditions in the training. 

\label{sec:data-proc}

\label{sec:datasets}
We aggregated IL property data from four publicly available repositories: ILTransR \cite{D3DD00040K}, Ionic Liquid Properties \cite{data4020088}, RBVAE-ANN-PSO \cite{CHEN2025335}, ILThermo \cite{Dong2007}. These sources collectively contributed 46,000 CO$_2$ solubility records, with 10k–16k entries per source and 23,000 viscosity records. 



We cleaned the data by removing noisy or inconsistent entries, data points recorded under extreme conditions, and records that violated basic physical consistency checks. The remaining entries were grouped by ionic liquid, temperature, and pressure—$(\mathrm{IL}, T, P)$—and averaged when intra-group variation fell within acceptable thresholds. After cleaning, the coverage of the data is described in Table \ref{tables/data-summary}. Data curation and processing details are summarized below. 

\subsubsection{Data curation and processing details}

We curated the combined dataset with three filters designed to (i) reduce duplicate/noisy measurements, (ii) align the training distribution with refinery-relevant operating envelopes, and (iii) enforce basic physical plausibility.

\begin{itemize}
  \item \textbf{Duplicate removal.} For entries sharing the same (cation, anion, $T$, $P$), we compared reported property values and discarded points whose relative deviation exceeded 5\%.
  \item \textbf{Extreme-condition exclusion.} To keep the dataset consistent with industrially relevant conditions, we removed measurements outside $T \le 420$ K and $P \le 200$ bar.
  \item \textbf{Physical consistency checks.} For solubility, we removed entries that violated expected monotonic trends (e.g., increasing solubility with increasing temperature at fixed pressure, or decreasing solubility with increasing pressure at fixed temperature). For viscosity, we removed entries exhibiting increasing viscosity with temperature (excluding cases plausibly explained by phase transitions).
\end{itemize}


\input{tables/data-summary}

For model and hyperparameter selection, and to ensure robustness of the performance estimates, we conducted a 5-fold cross-validation by creating balanced scaffold splits on Bemis-Murcko scaffolds applied to the cation of each ionic liquid in the dataset. This setup guarantees that each molecular family contributes to the final reported metrics and mitigates the risk of overfitting to specific molecular scaffolds, and preserves scaffold separation and leakage based on structure.

 Models are trained using the mean squared error (MSE) loss function on continuous targets, and evaluated using mean absolute error (MAE) and the coefficient of determination ($R^2$). These metrics are reported on the held-out scaffold-split test set across all experiments in Section \ref{sec:results}.

For our final D-MPNN predictions, we trained ensembles of 8 bootstrapped models with replacement for both predictive tasks. Each model was trained on a random subset of 90\% of the full dataset, and the randomly dropped-out data served as a test set for sanity checks in the style of bootstrap / out-of-bag validation. We trained on the full dataset for maximum predictive accuracy in an unknown chemical space.

\subsubsection{Baselines} To contextualize D-MPNN performance, we evaluated fingerprint-based gradient-boosted decision tree regressors such as XGBoost and CatBoost,  and a pre-trained ILTransR fine-tuning baseline. For the tree baselines, features include thermodynamic conditions  concatenated with Morgan fingerprints (radius 2) computed separately for cation and anion and then concatenated; 256-bit ion fingerprints are folded to 64 bits per ion via OR/max-pooling within 4-bit blocks. For ILTransR, we fine-tune the pretrained encoder with a small MLP regression head that maps the learned IL embedding concatenated with the condition vector to a scalar prediction. Head hyperparameters (dropout) are selected using the same 5-fold balanced cation-scaffold CV, choosing the configuration with the best mean held-out MAE ($R^2$ reported as a companion metric).

\subsection{Physics-Based Thermodynamic Property Estimation}
\label{sec:thermo}
The performance of IL solvents for post-combustion CO$_2$ capture depends on their operational absorption capacity and regeneration energy requirement. To quantify these downstream thermodynamic parameters, we perform post-processing of model-predicted CO$_2$ solubilities using standard physical approximations, enabling downstream ranking and feasibility filtering across the full IL candidate set.

Let $C(T)$ denote the predicted molar solubility of CO$_2$ at temperature $T$. For each candidate IL, we predict temperature-dependent CO$_2$ solubility at 20 evenly spaced temperature points in the range
$T \in [313.15\,\text{K},\,393.15\,\text{K}]$. All solubility predictions are made at a constant pressure of 1 bar, representative of flue gas conditions at standard refinery stack pressures.

It has been argued that the performance of a solvent in real applications cannot be reliably assessed from the equilibrium CO$_2$ absorption capacity, motivating the need to measure the difference in CO$_2$ solubility under adsorption and desorption conditions. This is often defined as the \textit{working capacity}: \cite{Raganati2020}

\begin{equation}
    \Delta C = C(T_\text{abs}) - C(T_\text{des})
\end{equation}

where $T_\text{abs} = 313.15\text{K}$ (40$^{\circ}$C) and $T_\text{des} = 393.15\text{K}$ (120$^{\circ}$C) are chosen to reflect realistic operating conditions for absorption and regeneration, respectively. This quantity describes the usable CO$_2$ capacity of each solvent, excluding reversible physisorption that may occur at elevated temperatures.

The heat of absorption, $\Delta H_{\text{abs, CO}_2}$, not only describes the heat released during absorption but also represents the intrinsic enthalpic cost of breaking the interaction between CO$_2$ and the solvent during regeneration. While not a direct measurement of energy demand, this value serves as an important parameter that strongly correlates with solvent regeneration costs.

In practical CO$_2$ capture processes, the total regeneration energy ($Q_{\text{reg}}$) involves multiple contributions beyond the desorption enthalpy alone \cite{Vinjarapu2024}. This can be expressed as:

\begin{equation}
Q_{\text{reg}} = Q_{\text{sens}} + Q_{\text{vap,H}_2\text{O}} + \Delta H_{\text{abs,CO}_2}
\end{equation}

where $Q_{\text{sens}}$ is the sensible heat needed to raise the solvent temperature to regeneration conditions and $Q_{\text{vap,H}_2\text{O}}$ is the latent heat required to generate stripping steam, typically when operating with aqueous diluted systems. While this decomposition highlights that $\Delta H_{\text{abs,CO}_2}$ is only a component of the total energy balance, it still offers a useful proxy for screening and ranking solvents. Solvents with lower $\Delta H_{\text{abs,CO}_2}$ generally require less energy for desorption and more favorable for reducing operational expenditure (OPEX) costs. 

To estimate the heat of absorption for each IL, we fit the predicted solubility data to a Van’t Hoff equation:
\begin{equation}
    \ln C(T) = -\frac{\Delta H_{\text{abs,CO}_2}}{R T} + b
\end{equation}

where $R = 8.3145 \times 10^{-3}$ kJ/mol·K is the universal gas constant, and $b$ is a linear intercept term.
For each IL, the slope of this linear regression across the 20 temperature points gives the estimated heat of absorption:
\begin{equation}
    \Delta H_{\text{abs,CO}_2} = -R \cdot \text{slope}
\end{equation}

We additionally compute the coefficient of determination $R^2$ for the linear fit to quantify goodness-of-fit. ILs with poor Van’t Hoff linearity (e.g., $R^2 < 0.8$) are penalized in subsequent scoring, as they may exhibit unpredictable regeneration behavior.

Accordingly, we use $\Delta H_{\text{abs,CO}_2}$ as a lower-bound estimate in our screening workflow to prioritize IL candidates with potential for low-energy regeneration. We can also directly compare the energy demand relative to traditional solvents such as MEA, which typically require 80–100 kJ/mol CO$_2$ \cite{Rochelle2009, Vega2016}.

To validate the reliability of our Van’t Hoff-based estimation framework, we benchmarked predicted $\Delta H_{\text{abs,CO}_2}$ values against 19 ILs with experimentally reported heats of absorption. These include ILs based on imidazolium, pyrrolidinium, and ammonium cations with a range of anions (e.g., [Tf$_2$N]$^-$, acetate, BF$_4$$^-$). The experimental values span from 10 kJ/mol (physisorbing ILs) to over 60 kJ/mol (chemisorbing ILs), in agreement with the literature \cite{Zhang2024}.

\subsection{Pareto-Based Candidate Selection}
\label{sec:pareto-selection}

Selecting optimal ILs for CO$_2$ capture requires balancing multiple competing objectives: absorption capacity, transport properties, and regeneration efficiency. Rather than collapsing these trade-offs into a single weighted score, we adopt a multi-objective optimization strategy to find the best set of ILs that achieve a satisfactory balance among all the criteria---the \textit{Pareto front}. This approach identifies candidates that offer the best trade-offs, where improving one objective would necessarily worsen another.

To ensure data quality and physical plausibility, we first filter candidates before Pareto optimizing using the criteria listed in Table~\ref{tab:filtering_criteria}

\input{tables/filtering-criteria}

We then apply fast non-dominated sorting via the \texttt{paretoset} Python package, which implements the NSGA-II algorithm, to identify candidates on the Pareto frontier \cite{Odgaard2021, Deb2002}. The objectives include maximizing the working capacity ($\Delta C$), maximizing the CO$_2$ solubility at scrubbing conditions ($x_{\mathrm{CO_2, scrub}}$), and minimizing the viscosity ($\ln \eta$). The resulting non-dominated set (Tier 1) consists of candidates where no property can be improved without worsening another. Lower tiers represent increasingly dominated solutions. Within each tier, we optionally compute Euclidean distance to the Pareto frontier as a tie-breaking metric.

This method avoids arbitrary weighting and reflects best practices from multi-criteria decision-making (MCDM) in materials discovery \cite{Andre2023, Fromer2024}. It allows interpretable exploration of trade-offs between process-relevant properties and supports flexible down-selection of ILs for experimental follow-up.

\subsection{Synthesis Feasibility Screening}
\label{methods:synthesis-feas}

The final filter used to select IL candidates accounts for the viability of laboratory and industrial synthesis pathways. We apply this analysis to the top candidates from the Pareto evaluation step outlined in Section \ref{sec:pareto-selection}. Retrosynthetic analysis was performed using the ASKCOS platform, enhanced with Monte Carlo Tree Search (MCTS) to efficiently explore synthetic pathways. To quantify the synthetic feasibility of a route, we computed and reported the average plausibility across all steps in the retrosynthetic tree, the average depth of trees found per IL, and the highest plausibility value across all retrosynthetic trees for a given IL. 



By incorporating synthesis feasibility screening into our pipeline, we ensure that the final shortlist of ILs reflects both chemical promise and manufacturability.

%% file: tables/data-summary.tex
\begin{table}[t]
  \setlength{\tabcolsep}{13.5pt}
  \centering
  \caption{Summary of IL property data collected for model training.}
  \begin{tabular}{lrr}
    \toprule
    \multirow{2}{*}{\textbf{Dataset details}} &
      \multicolumn{2}{c}{\textbf{Property}} \\           
    \cmidrule(lr){2-3}
     & \textbf{CO\textsubscript{2} solubility} & \textbf{Viscosity} \\  
    \midrule
    Total records            & 17,261 & 20,309 \\
    Unique ILs               &     414 &  2,744 \\
    Unique anions            &     107 &    315 \\
    Unique cations           &     159 &  1,218 \\
    Temperature range (K)    & 243--420 & 253--420 \\
    Pressure range (bar)     &   0--200 & -- \\
    Target property range    & $x_{\mathrm{CO_2}}$: 0--1 mol CO$_{2}$ mol$^-1$ & 0.74--$7.08\times10^{6}$ mPa $\cdot$ s \\
    \midrule
    Possible ion combinations         & 17,013   &   383,670   \\
    \bottomrule
  \end{tabular}
  \label{tables/data-summary}
\end{table}

%% file: tables/filtering-criteria.tex
\begin{table}[t]
\setlength{\tabcolsep}{32pt}
\centering
\caption{Filtering criteria applied prior to Pareto front analysis to ensure physical plausibility and process relevance.}
\label{tab:filtering_criteria}
\begin{tabular}{ll}
\toprule
\textbf{Criterion} & \textbf{Threshold} \\
\midrule
Working capacity, $\Delta C$ & $> 0.10$ mol CO$_2$/total mols \\
Natural log viscosity, $\ln \eta$ & $< 4.6$ (approx. $\eta < 100$ mPa·s) \\
Regeneration energy, $-\Delta H_{\mathrm{abs,CO_2}}$ & $> 1$ kJ/mol CO$_2$ \\
\bottomrule
\end{tabular}
\end{table}

%% file: sections/03_results.tex
\section{Results and Discussion}
\label{sec:results}

We present the results of our AI-guided pipeline for identifying and ranking ionic liquid (IL) candidates under typical refinery flue gas CO$_2$ capture conditions. We focus our discussion on three complementary results: (i) model predictive performance, (ii) analysis of desired IL properties for effective and practical carbon capture, and (iii) identification of top-performing candidates.

\subsection{Model Performance and Predictive Confidence}
\label{sec:results-model}

We begin by evaluating the robustness and generalizability of our GNN-based models for predicting IL properties. Both D-MPNN models achieved high predictive accuracy, with average $R^2$ scores of $0.89$ and $0.61$ each respectively. The average mean absolute errors (MAE) recorded while evaluating the model were $0.18$ for solubility and $0.53$ log-units for viscosity. These metrics were calculated on held-out scaffold-split test sets. On the bootstrapped out-of-bag test evaluations (random and not chemistry-aware),  both models record R$^2$ scores well over $0.95$ and less than half of the above MAEs.

{\textbf{Baseline comparison.} Under the same balanced cation-scaffold split protocol, we evaluated XGBoost, CatBoost, and ILTransR fine-tuning baselines. Table~\ref{tab:baseline_perf} summarizes CV (pooled out-of-fold) and scaffold-disjoint test performance across both targets.

{\textbf{Distribution comparison.} We compare the property distributions with kernel density estimation plots in Figure \ref{fig:kpe-dists} to determine how well the D-MPNN models the training data. The predicted property distributions of the generated ILs show significant overlap with those from the training datasets for CO$_2$ solubility and viscosity indicating good predictive performance. Additionally, the distributions of the Pareto front candidates exhibit the shift towards desired properties: higher CO$_2$ solubility and lower viscosity respectively.

\begin{figure}[t]
    \centering
    \includegraphics[width=0.85\linewidth]{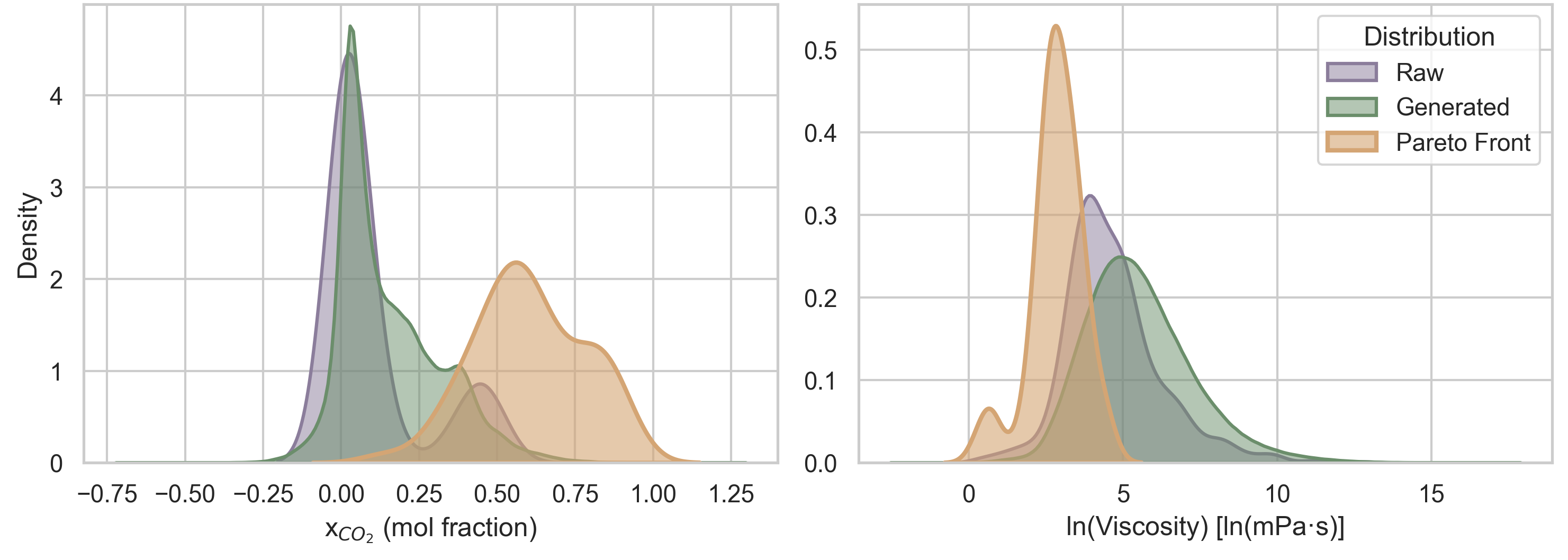}
    \caption{Distributions of predicted CO$_2$ solubility and natural log viscosity properties for generated ionic liquids compared to those from raw training dataset. Plots show property distributions at absorption conditions ($T=313.15 \text{K}, P=1 \text{bar}$). Pareto front distribution represent isolated top performers and a subset of generated.}
    \label{fig:kpe-dists}
\end{figure}

\begin{table}[t]
\setlength{\tabcolsep}{7pt}
\centering
\caption{Baseline performance under 5-fold balanced cation-scaffold CV and scaffold-disjoint held-out test evaluation. CV metrics are computed from pooled out-of-fold predictions across the five folds.}
\label{tab:baseline_perf}
\begin{tabular}{ll l r r r r}
\toprule
Task & Target & Model & CV MAE & CV $R^2$ & Test MAE & Test $R^2$  \\
\midrule
Viscosity & $\ln(\eta)$ & XGBoost & 0.895 & 0.512 & 0.659 & 0.635   \\
Viscosity & $\ln(\eta)$ & CatBoost & 0.914  & 0.500 & 0.709 & 0.599  \\
Viscosity & $\ln(\eta)$ & ILTransR (finetune) & 0.864 & 0.441 & 0.670 & 0.608  \\
CO$_2$ solubility & $x_{\mathrm{CO_2}}$ & XGBoost & 0.086 & 0.660 & 0.078 & 0.779 \\
CO$_2$ solubility & $x_{\mathrm{CO_2}}$ & CatBoost & 0.085 & 0.705 & 0.080 & 0.808 \\
CO$_2$ solubility & $x_{\mathrm{CO_2}}$ & ILTransR (finetune) & 0.110 & 0.613 & 0.094 & 0.755  \\
\bottomrule
\end{tabular}
\end{table}

Since we trained 8 bootstrapped models for prediction, we were able to study the spread in ensemble predictions to determine the uncertainty of predictions. We quantify prediction uncertainty via ensemble agreement diagnostics, summarized below.

\subsubsection{Ensemble agreement and prediction uncertainty}

To assess agreement between the eight bootstrapped ensemble models, we compute pairwise comparisons across model pairs. For CO$_2$ solubility, we evaluate agreement across all ionic-liquid/temperature combinations (IL$\times T$) by plotting $y_j$ (model $j$) versus $y_i$ (model $i$) with the identity line $y=x$ indicating perfect agreement. Within each pairwise panel we report: Pearson correlation ($r$), root mean squared error (RMSE), mean bias ($y_j-y_i$), and Bland--Altman limits of agreement (LoA), reported as half-width $\pm 1.96\,\sigma_d$ where $d_k=y_{j,k}-y_{i,k}$ and $\sigma_d$ is the standard deviation over rows $k$. Each point corresponds to the same IL at the same temperature for both models, so temperature is controlled at the point level while the grid pools all temperatures.

\begin{figure}[!h]
    \centering
    \includegraphics[width=0.85\linewidth]{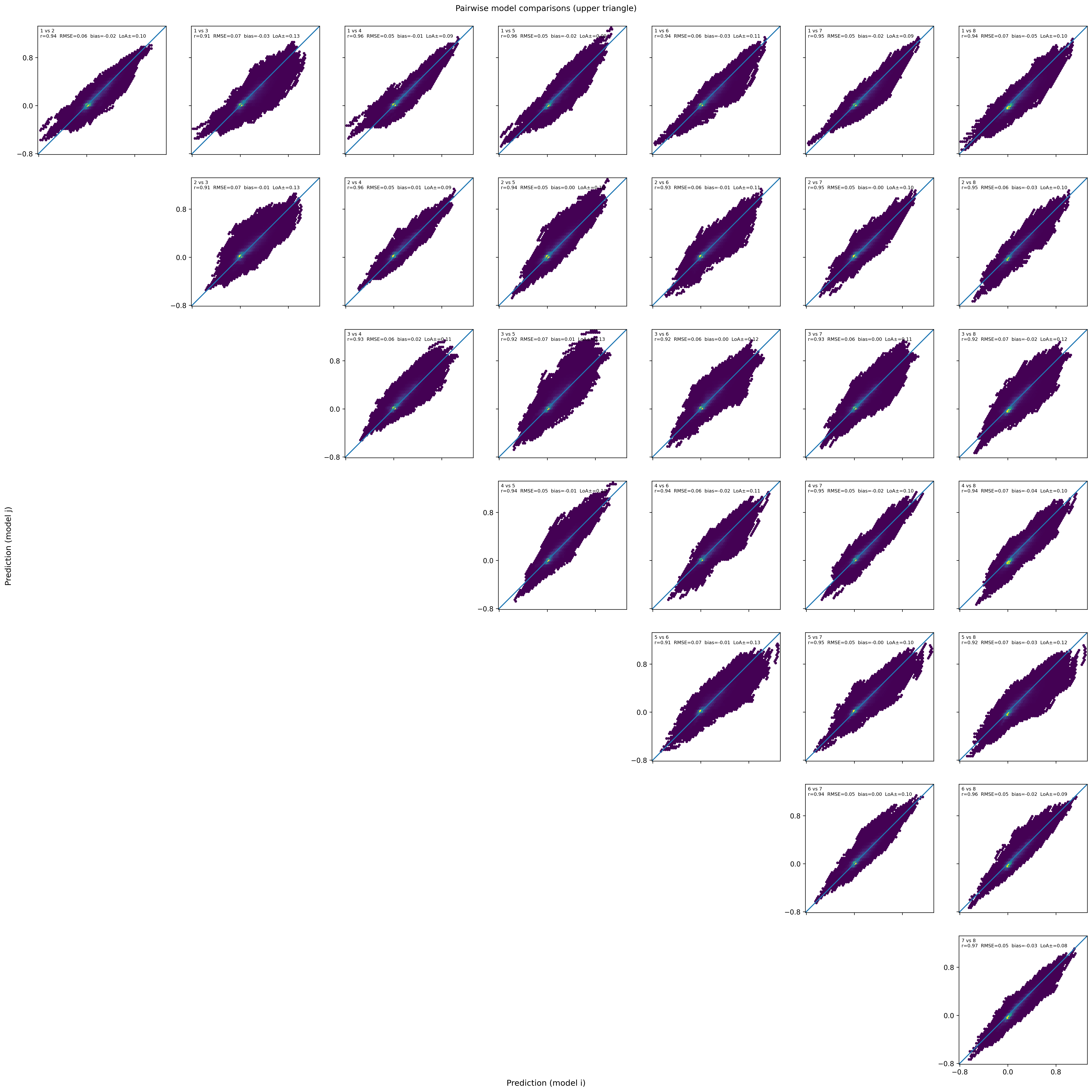}
    \caption{Pairwise model agreement for CO$_2$ solubility. Hexbin plots (model $j$ vs.\ $i$) across all IL$\times$T with $y{=}x$; panels report $r$, RMSE, bias ($y_j{-}y_i$), and LoA half–width ($\pm 1.96\,\sigma_d$)}
    \label{fig:co2_uncertainty}
\end{figure}

To reduce overplotting at scale, we visualize density via hexagonal binning and keep a fixed global axis range across panels to enable direct comparison of offsets and dispersion. For viscosity (predicted in log units), we perform the analogous pairwise comparisons; because predictions are in logarithmic units, bias and LoA admit a multiplicative interpretation via $\exp(\cdot)$.

\begin{figure}[!h]
    \centering
    \includegraphics[width=0.85\linewidth]{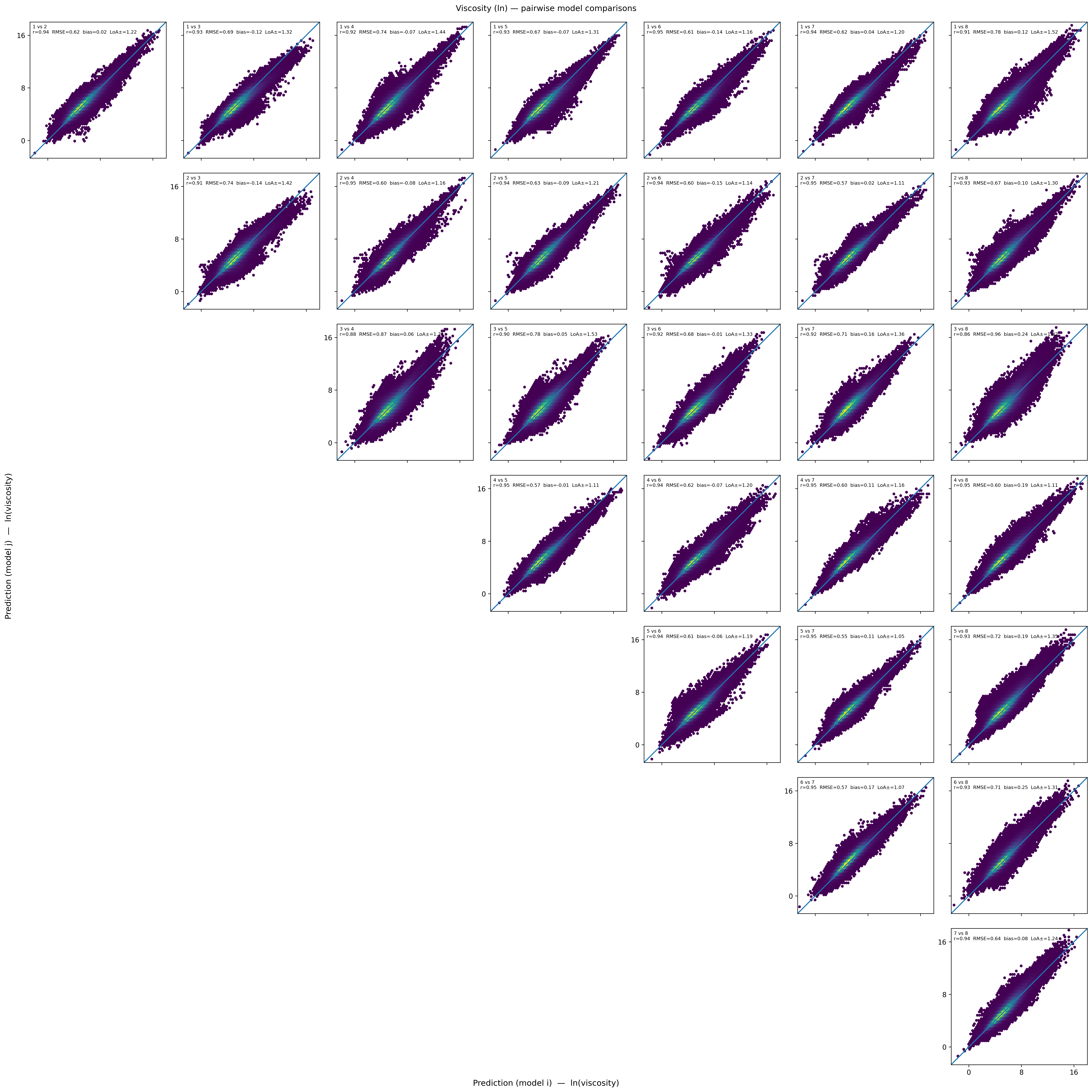}
    \caption{Pairwise model agreement for $\ln(\eta)$. Hexbin plots (model $j$ vs.\ $i$) across all IL$\times$T with $y{=}x$; panels report $r$, RMSE, bias ($y_j{-}y_i$), and LoA half–width ($\pm 1.96\,\sigma_d$)}
    \label{fig:viscosity_uncertainty}
\end{figure}

Finally, we summarize ensemble spread for selected candidates with box-and-whisker plots. For viscosity, we plot the eight-model log-viscosity predictions at $T = 313.15 K$ for the 20 ILs with the highest across-model mean (IL 1–IL 20, anonymized) in Figure \ref{fig:viscosity_uncertainty_candidates}. For solubility, for each temperature we select the ten ILs with the highest across-model mean predicted solubility and show the distribution of the eight ensemble predictions (IL 1–10 per temperature panel, anonymized) in Figure \ref{fig:co2_candidate_uncertainty}. Narrow boxes indicate stronger agreement among models (lower epistemic uncertainty), whereas wider boxes and outliers indicate greater model disagreement.

\begin{figure}[!h]
    \centering
    \includegraphics[width=0.80\linewidth]{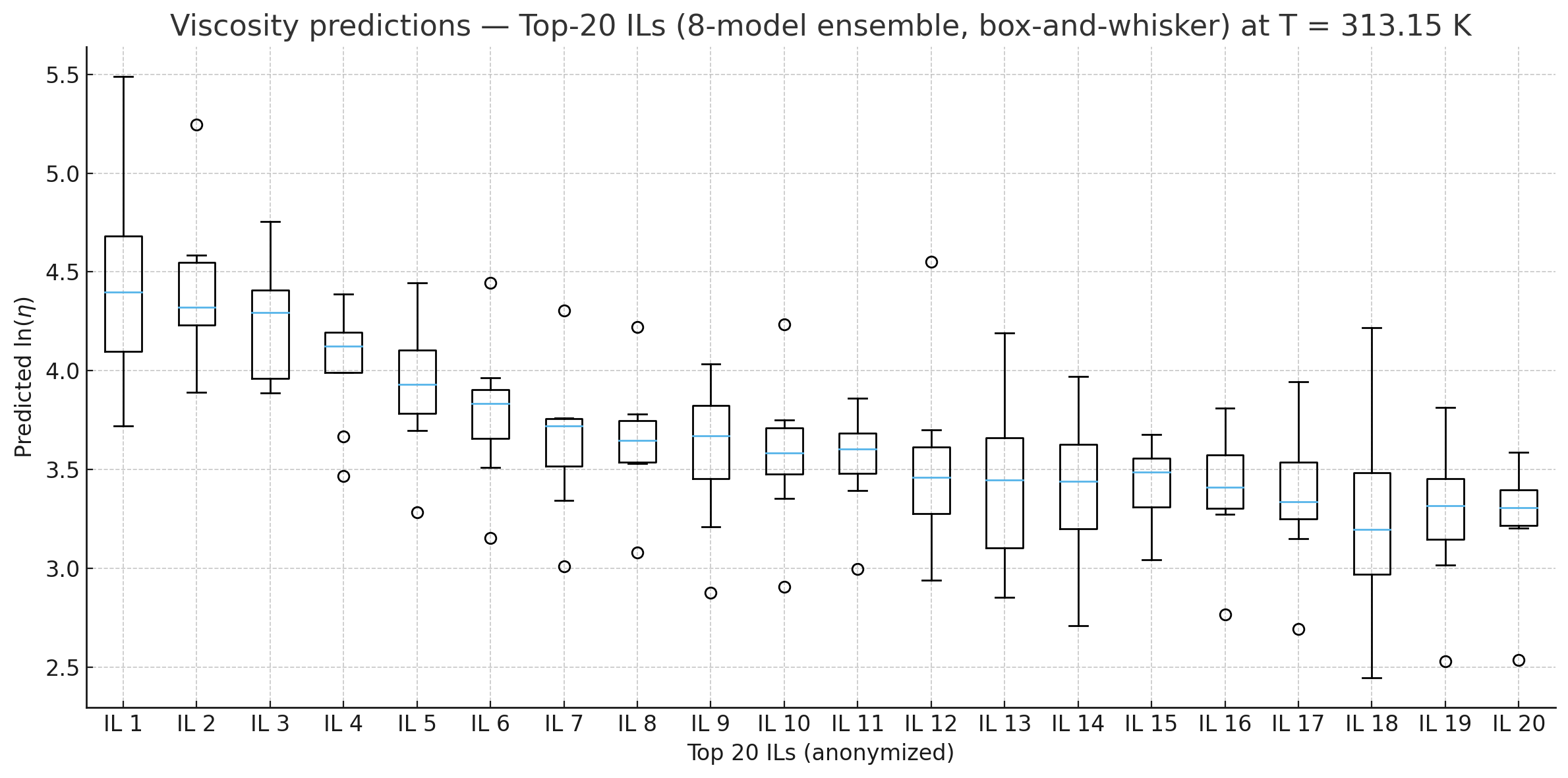}
    \caption{Box-and-whisker summaries of eight-model viscosity predictions at 313.15 K for the 20 highest-mean ILs (IL 1–IL 20), showing median, IQR, 1.5×IQR whiskers, and outliers}
    \label{fig:viscosity_uncertainty_candidates}
\end{figure}

\begin{figure}[!h]
    \centering
    \includegraphics[width=1\linewidth]{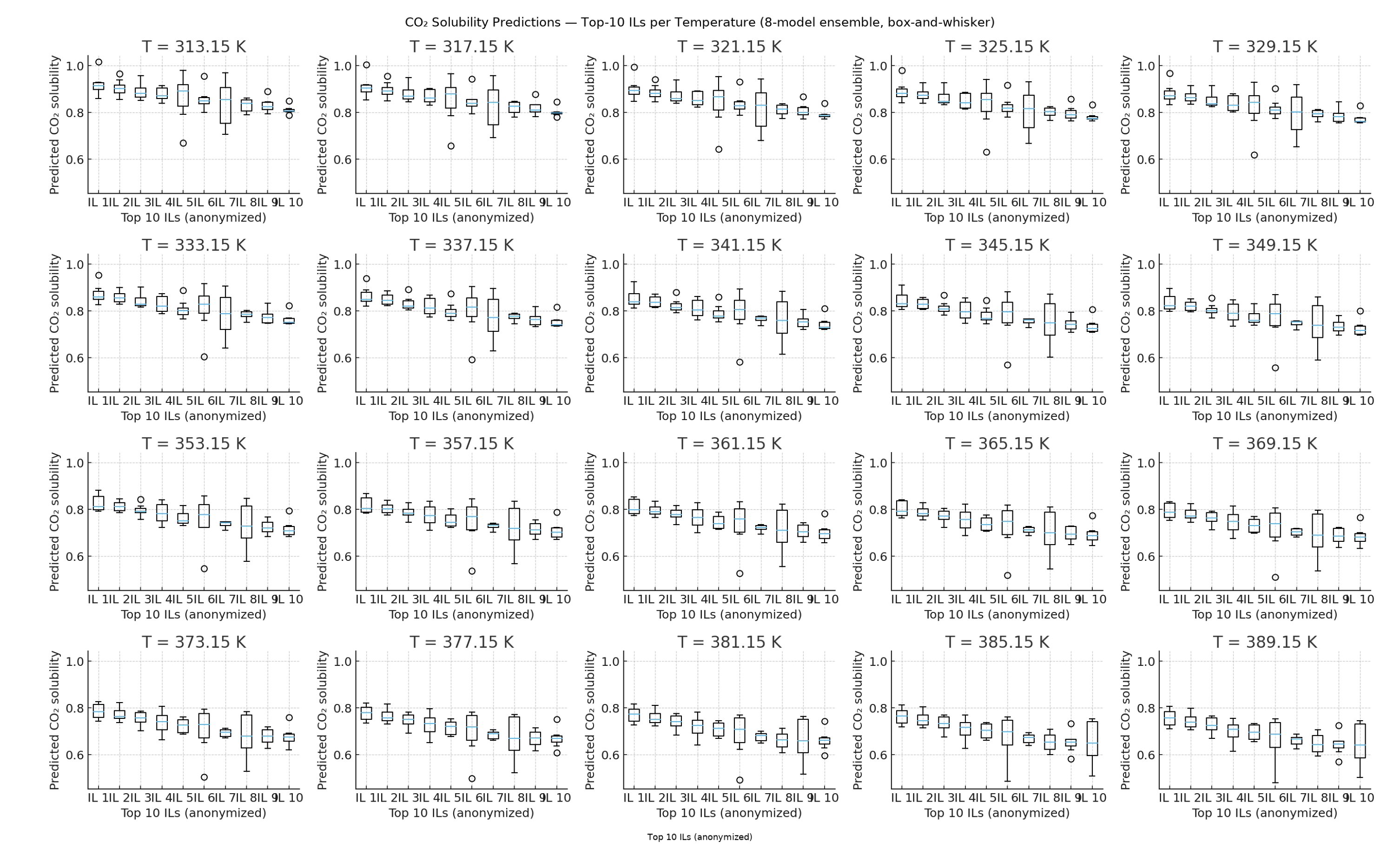}
    \caption{Ensemble-prediction uncertainty for CO$_{2}$ solubility: per-temperature box-and-whisker plots for the top-10 ILs (ranked by 8-model mean; ILs anonymized), with a common y-axis for cross-temperature comparison}
    \label{fig:co2_candidate_uncertainty}
\end{figure}

\subsection{Desired Property Analysis: Working Capacity, Max-loading Viscosity, and Regeneration Energy Estimation}

\begin{figure}[t]
    \centering
    \includegraphics[width=0.85\linewidth]{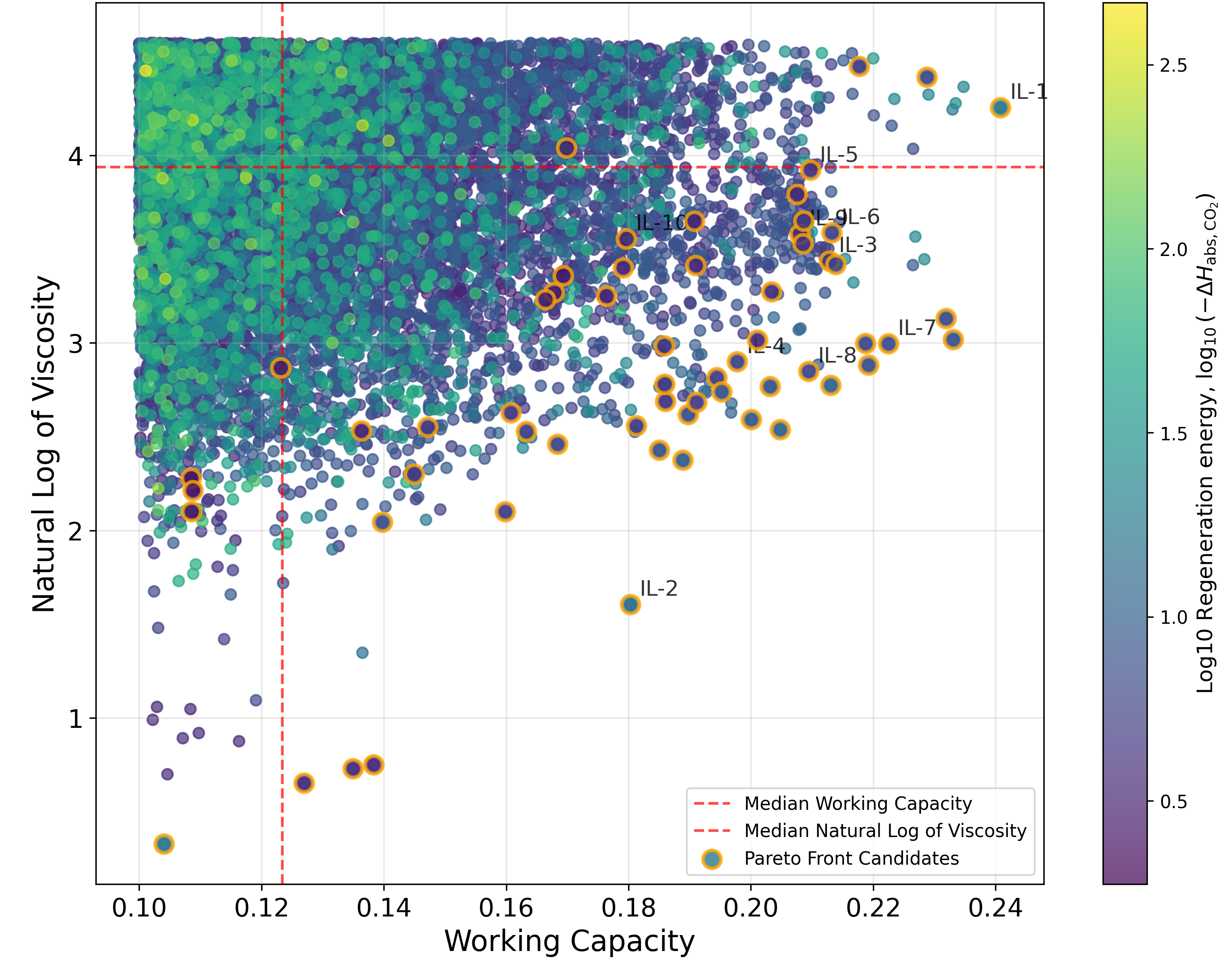}
    \caption{Working capacity versus viscosity for filtered IL candidates, colored by log-scale heat of absorption with Pareto-front candidates highlighted. High-performing ILs cluster in the lower-right quadrant, indicating favorable capacity, fluidity, and energy demand}
    \label{fig:work-cap-visc}
\end{figure}

We define working capacity as the difference in predicted CO$_2$ loading at 40$\degree$C and 120$\degree$C at 1 bar, simulating absorption and regeneration conditions. Figure~\ref{fig:work-cap-visc} plots the predicted natural log of the max-loading viscosity against working capacity. The ideal candidates appear in the bottom-right quadrant with desirable properties---high capacity and low viscosity.

\subsubsection{Coupling between maximum CO$_2$ loading and viscosity}
A key practical constraint is that strong absorption at high loading can coincide with prohibitively high viscosity. Figure~\ref{fig:loading_vs_viscosity} shows maximum predicted CO$_2$ loading versus predicted viscosity and highlights the nonlinear coupling between these quantities. Many candidates with favorable loading are excluded by rich-phase viscosity, reinforcing the need for viscosity-aware screening; relatively few high-loading candidates remain below $\sim$100 mPa$\cdot$s, a commonly used operational threshold.



To estimate the thermal energy cost of regeneration, we fit predicted temperature-dependent solubility curves for each IL to a Van't Hoff linear model. The resulting slope yields the enthalpy of absorption ($\Delta H_{\text{abs,CO}_2}$), while R$^2$ provides a proxy for fit quality. Across all IL candidates, the average $\Delta H_{\text{abs,CO}_2}$ was $-26.5 \pm 41.2$ kJ/mol. As seen in Figure \ref{fig:heat-results}, the majority of high-performing ILs showed good linearity (mean $R^2$ > 0.90), suggesting physically plausible and temperature-sensitive behavior. The candidates with the best fits show $\Delta H_{\text{abs,CO}_2}$ in the range of 5–20 kJ/mol.

\subsubsection{Example Van’t Hoff fits}

We observe a range in Van’t Hoff fit quality across candidates. Low-quality fits may indicate that a constant-enthalpy approximation is insufficient over the evaluated temperature range, and can motivate higher-order (e.g., polynomial) fits for candidates exhibiting nonlinearity. Representative examples are shown in Figure~\ref{fig:vant-hoff-plots}.


To contextualize the performance of newly identified candidates, we benchmarked our predictive models against a curated set of previously reported ILs for CO$_2$ heat of absorption. Figure \ref{fig:heat-results} summarizes the comparison of predicted values against these benchmark ILs. While absolute prediction error varies across ILs, we find that the general order of magnitude and relative rank orderings of predicted $\Delta H_{\text{abs,CO}_2}$ are broadly consistent with known literature values.

\begin{figure}[t]
    \centering
    \begin{minipage}[t]{0.39\linewidth}
        \centering
        \includegraphics[width=\linewidth]{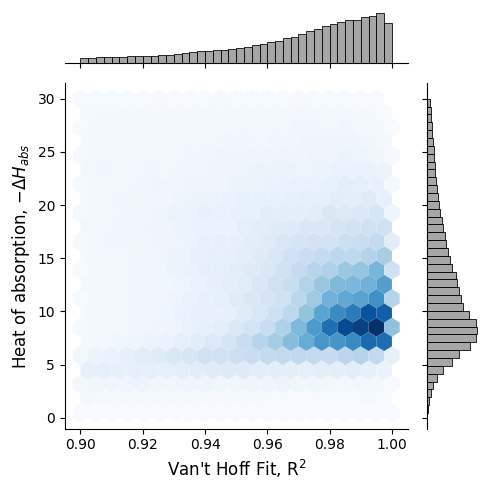}
    \end{minipage}
    \hfill
    \begin{minipage}[t]{0.60\linewidth}
        \centering
        \includegraphics[width=\linewidth]{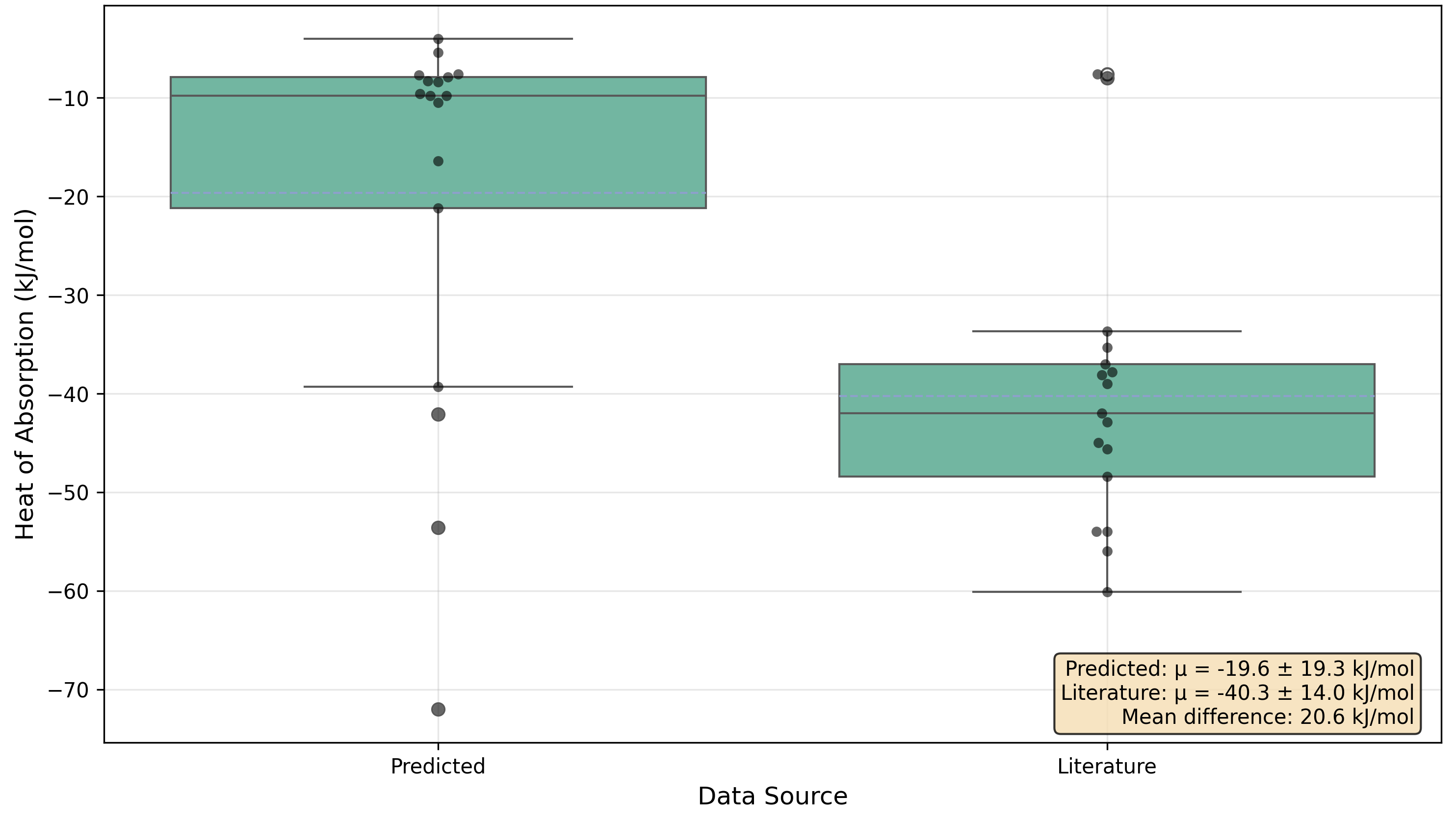}
    \end{minipage}
\caption{\textbf{Left.} Distribution of predicted heat of absorption (-$\Delta H_{\text{abs,CO}_2}$) versus Van’t Hoff fit ($R^2$). \textbf{Right.} Heat of absorption benchmarking. Predicted IL $-\Delta H_{\text{abs,CO}_2}$ values are generally consistent with literature ranges, though slightly lower.}
\label{fig:heat-results}
\end{figure}

\input{tables/benchmark-heats}


\subsection{Family-Level Insights and Chemical Trends}
To better understand the structure-property relationships of high-performing ILs, we grouped candidates by their cation and anion family labels (e.g., imidazolium, phosphonium, amino acid, sulfonate). Figure \ref{fig:family_visc}} aggregates working capacity property predictions for each family pair. Figure~\ref{fig:family_visc} shows the corresponding viscosity landscape (family-average log-viscosity; lower is better). Importantly, the observed high-scoring regions include not only combinations reported in literature but also novel pairings that fall outside of the experimentally characterized space, demonstrating the capacity of the predictive framework to identify unexplored ILs with promising predicted properties.

\begin{figure}[t]
    \centering
    \includegraphics[width=1.0\linewidth]{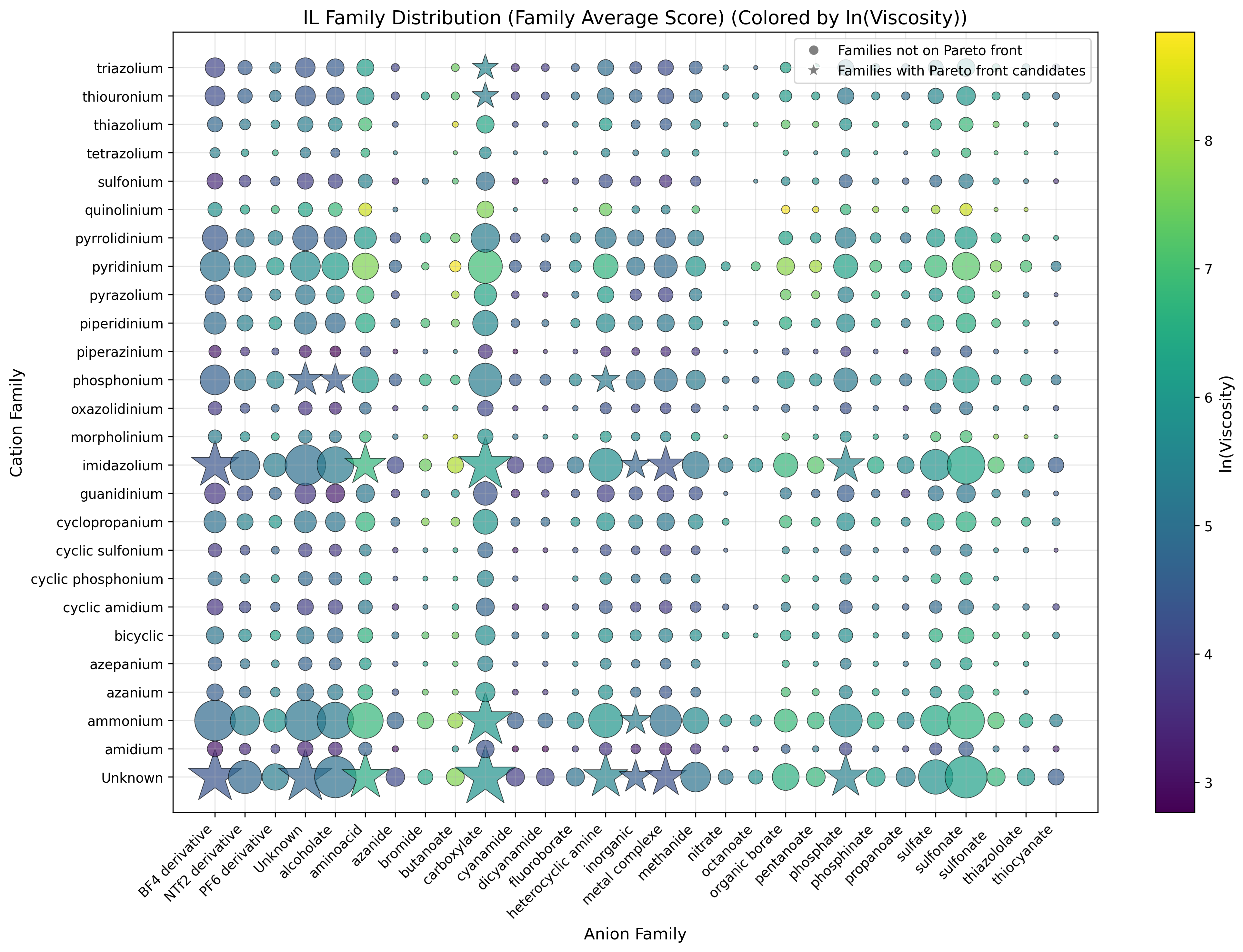}
    \caption{Bubble chart of IL candidates grouped by cation and anion families. Stars indicate family combinations with ILs represented in the Pareto front candidates. Bubble size is proportional to the number candidates. Color indicates average natural log viscosity across the family combination candidates, where lower is better}
    \label{fig:family_visc}
\end{figure}

\begin{figure}[t]
    \centering
    \includegraphics[width=1\linewidth]{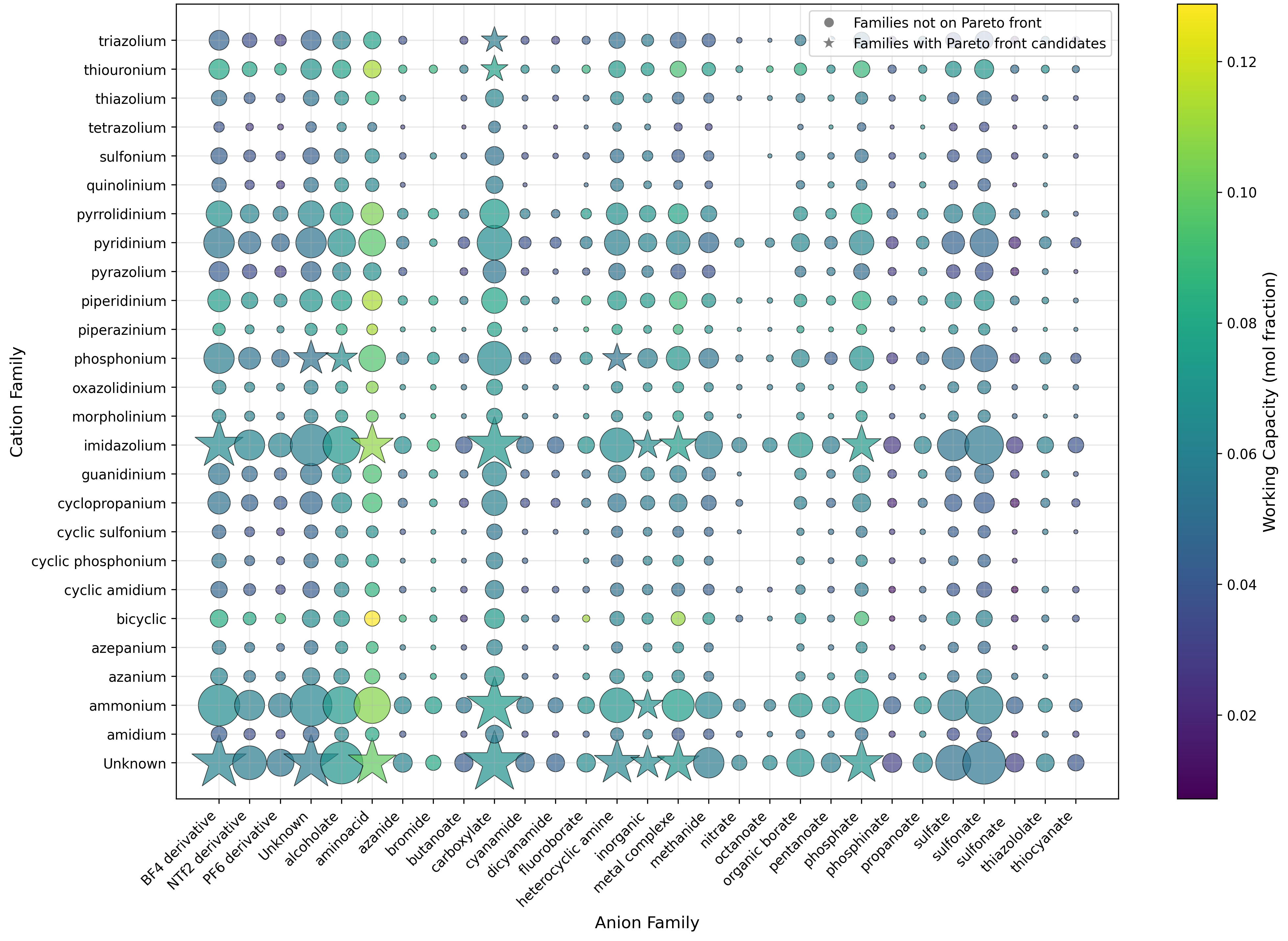}
    \caption{Bubble chart of IL candidates grouped by cation and anion families. Stars indicate family combinations with ILs represented in the Pareto front candidates. Bubble size is proportional to the number candidates. Color indicates average working capacity across the family combination candidates, where higher is better}
    \label{fig:family-performance}
\end{figure}

We see that amino acid anion-based anion ILs tend to consistently perform well in working capacity. Interestingly, imidazolium cation-based ILs, which are generally acknowledged as conventional ILs for CO$_2$ capture tended to represent a good proportion of top candidate ILs, showing strong alignment with literature \cite{Zafar2025}.

\begin{figure}[t]
    \centering
    \includegraphics[width=0.75\linewidth]{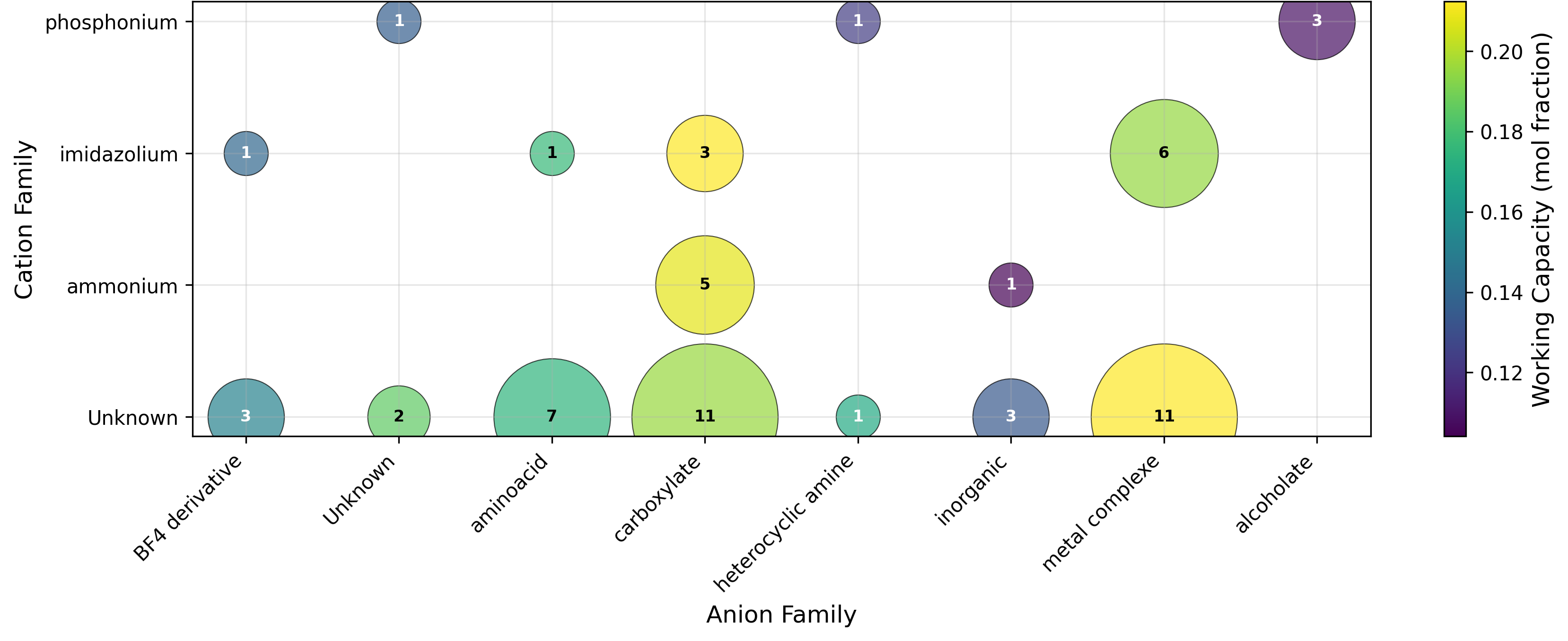}
    \caption{Bubble chart of Pareto front IL candidates grouped by cation and anion families. Bubble size is proportional to the number candidates. Color indicates average working capacity across the family combination candidates, where higher is better}
    \label{fig:pareto-family-performance}
\end{figure}

Figure \ref{fig:pareto-family-performance} shows a focused view of the working capacity trends by family combination for the Pareto front candidate ILs. We observe the majority of IL candidates with known families fall into the imidazolium cation-based, carboxylate anion-based, and metal complexe anion-based categories. Interestingly, many of the top performing IL cation families are unknown, requiring further analysis to infer any further chemical trends.

\subsection{Survey of Top-Performing Candidate ILs}
We performed retrosynthetic planning using the ASKCOS Tree Builder, which employs a template-based Monte Carlo Tree Search (MCTS) algorithm to explore synthetic pathways. The search was conducted in parallel across eight active pathways, allowing multiple branches of the retrosynthetic tree to be expanded simultaneously. At each node, up to 20 reaction templates were applied to generate possible precursor sets. The search proceeded up to a maximum depth of four reaction steps, with a total expansion time capped at 60 seconds. During this period, ASKCOS iteratively constructed and evaluated candidate routes, terminating the search upon identifying up to 500 retrosynthetic trees whose leaf nodes consisted entirely of commercially available starting materials. Each proposed reaction step was scored using a forward reaction prediction model, and for each complete route, we computed the average plausibility score by taking the mean of individual step scores. We averaged plausibility over all complete routes to come to a final plausibility score for any given ionic liquid, and also reported the maximum plausibility score across all synthesis routes. Over 70\% of top ILs were assigned synthesis routes. The results from the synthesis feasibility analysis are outline in Table \ref{tab:top_il_summary}. Note that the feasibility does not correlate in any way with the property performance, emphasizing the importance of having a late-stage synthesis filter to discover commercially viable candidates. 


\label{sec:results-top-ils}
Table \ref{tab:top_il_summary} reports a sample of anonymized ILs drawn from the Pareto front candidate ILs. For confidentiality reasons, structural identifiers are withheld. Instead, we present predicted values for the four criteria of interest: working capacity, regeneration energy, Van’t Hoff R², and viscosity at full absorption. These ILs are notable for achieving high capacity at moderate regeneration energy and manageable viscosity, with strong Van’t Hoff fits indicating thermodynamic regularity across the target temperature range. Several exhibit performance profiles that surpass even the best benchmarked ILs, highlighting the practical value of combining predictive modeling with synthesis-aware optimization.

In ongoing work, these candidates are undergoing additional screening for thermal and oxidative stability, environmental toxicity, and synthesis feasibility under industrial constraints.

\input{tables/top-candidates}

\subsection{Economic analysis (order-of-magnitude comparison)}

To contextualize process impact, Table \ref{tab:capex_opex} summarizes an order-of-magnitude comparison between conventional amines and ionic liquids across key cost levers reported in the literature (e.g., stability/loss, replacement cost, and regeneration energy). While ILs can carry higher upfront solvent cost, their reduced loss rates and lower regeneration energy can translate to lower operating cost ranges.

\input{tables/capex_opex}

%% file: tables/benchmark-heats.tex
\begin{table}[t]
\setlength{\tabcolsep}{7pt}
\centering
\small
\caption{Comparison of predicted and literature CO$_2$ absorption enthalpies for selected benchmark ILs.}
\label{tab:benchmark-heats}
\begin{tabular}{lrrrr}
\toprule
IL Abbreviation & Predicted & Literature & Absolute Error & Percent Error \\
                & $-\Delta \hat{H}_{\mathrm{abs,CO_2}}$ (kJ/mol) & $-\Delta H_{\mathrm{abs,CO_2}}$ (kJ/mol) &  &  \\
\hline
{[P4442][Ph-Suc]}  & 42.1 & 48.4 & 6.3  & 12.9\% \\
{[EMIM][Tf2N]}     & 72.0 & 54.0 & 18.0 & 33.3\% \\
{[Ipmim][Triz]}    & 9.6  & 37.8 & 28.2 & 74.6\% \\
\bottomrule
\end{tabular}
\end{table}

%% file: tables/top-candidates.tex
\begin{table}[t]
\centering
\small
\caption{Summary of top IL candidates with favorable absorption and regeneration properties, and synthesis feasibility analysis. ILs with asterisk are highlighted for high synthesis feasibility. All names anonymized for confidentiality.}
\label{tab:top_il_summary}
\renewcommand\arraystretch{1.2}
\begin{tabularx}{\textwidth}{l
  >{\centering\arraybackslash}X 
  >{\centering\arraybackslash}X 
  >{\centering\arraybackslash}X 
  >{\centering\arraybackslash}X 
  >{\centering\arraybackslash}X 
  >{\centering\arraybackslash}X 
  >{\centering\arraybackslash}X 
  >{\centering\arraybackslash}X 
  >{\centering\arraybackslash}X}
\toprule
\makecell{IL\\ID} & 
\makecell{Working\\Capacity\\(mol/mol)} & 
\makecell{Viscosity\\(mPa·s)} & 
\makecell{$-\Delta H_{\text{abs}}$\\(kJ/mol CO$_2$)} & 
\makecell{Van’t\\Hoff $R^2$} & 
\makecell{Synthetic\\Trees \\Found} & 
\makecell{Avg.\\Tree \\Depth} & 
\makecell{Avg.\\Tree\\Plausibility} & 
\makecell{Max.\\Plausibility\\Tree} \\
\midrule
IL-1 & 0.241 & 70.37 & 17.52 & 0.958 & 0   & --   & --   & --    \\
\textbf{IL-2*} & 0.180 & 4.97 & 13.17 & 0.998 & 51   & 3.510 & 0.265 & 0.637  \\
IL-3 & 0.213 & 31.05 & 4.52 & 0.992 & 83   & 4.000 & 0.159 & 0.477  \\
IL-4 & 0.198 & 18.15 & 5.72 & 0.992 & 140   & 3.943 & 0.163 & 0.492  \\
IL-5 & 0.210 & 50.42 & 4.22 & 0.992 & 1 & 1.000 & 0.002 & 0.002  \\
\textbf{IL-6*} & 0.213 & 36.11 & 5.69 & 0.989 & 54  & 3.352 & 0.393 & 0.628 \\
IL-7 & 0.222 & 19.99 & 6.61 & 0.997 & 0 & -- & -- & --  \\
IL-8 & 0.210 & 17.25 & 6.10 & 0.998 & 0   & --   & --   & --    \\
IL-9 & 0.208 & 35.82 & 3.63 & 0.993 & 2   & 2.000 & 0.027 & 0.052  \\
\textbf{IL-10*} & 0.180 & 34.95 & 3.12 & 0.995 & 1   & 2.000 & 0.860 & 0.860  \\
\bottomrule
\end{tabularx}
\end{table}

%% file: tables/capex_opex.tex
\begin{table}[t]
\setlength{\tabcolsep}{17.5pt}
\caption{Despite higher production costs, ILs show economic advantages over amines across several key cost levers. For each critical factor assessed to compare amines and ILs, the superior performer is bolded.}
\centering
\begin{tabular}{@{}lcc@{}}
\toprule
\textbf{Critical factor to assess} & \textbf{Amines} & \textbf{Ionic Liquids} \\ \midrule
CO$_2$ loading capacity          & \textbf{0.36 (ton CO$_2$/ton) }           & 0.20 (ton CO$_2$/ton) \\
Solvent stability/loss           & 120--250 (g/ton CO$_2$)          & \textbf{10--20 (g/ton CO$_2$)} \\
Solvent cost                     & \textbf{1{,}000--2{,}000 (\$/ton)}        & 3{,}000--5{,}000 (\$/ton) \\
Solvent replacement cost         & 0.19--1.31 (\$/ton CO$_2$)       & \textbf{0.01--0.04 (\$/ton CO$_2$)} \\
Regeneration energy              & 3.5 (GJ/ton CO$_2$)              &\textbf{ 2 (GJ/ton CO$_2$)   }   \\ \midrule \midrule
\textbf{Estimated OPEX cost (\$/ton CO$_2$)} & \textbf{20--40} & \textbf{10--20} \\ \bottomrule
\end{tabular}
\label{tab:capex_opex}
\end{table}

%% file: sections/99_conclusions.tex
\section{Conclusion}
\label{sec:conclusions}
This work presents a tightly integrated and fully automated modular, scalable, and interpretable framework for the data-driven discovery of ionic liquid (IL) solvents for CO$_2$ capture under industrial operating conditions. 


Through this novel framework, we generated over 400,000 candidate ILs and predicted desirable properties for each. By leveraging a multi-objective optimization strategy, we identified 60 of the highest performing IL candidates that are on the Pareto front, of which 36 are found to have viable synthesis routes. These candidates are predicted to show strong performance across desired properties. With competitive predicted CO${_2}$ working capacity, manageable viscosity, and significantly lower regeneration energy, these candidates may be viable replacements for traditional solvents like MEA and MDEA. We present the properties of ten proprietary candidates. These candidates deliver working capacities of 0.18–0.24, low regeneration enthalpies around 10 kJ mol$^{-1}$, and viscosities below 70 mPa$\cdot$s while maintaining Van’t Hoff linearity ($R^2 \geq$  0.95). Retrosynthetic analysis found viable synthetic routes for seven (70\%) of these molecules; three candidates had at least one synthetic route scoring over 0.62 in a forward-predicted plausibility score. The best performers consistently pair carboxylate or metal complex anions with imidazolium or ammonium cations, reinforcing literature trends and providing a clear starting points for experimental validation.

Although more laboratory and pilot plant validation of IL properties is needed, the possibility of deploying some of the identified candidates could lead to significant cost reductions for energy companies. Despite higher upfront solvent costs, the superior thermal stability and significantly lower regeneration energy of ILs contribute to estimated OPEX reductions of 5–10\% compared to conventional amines. Additionally, their non-volatile, corrosion-resistant nature could enable up to 10\% CAPEX savings for an enterprise decarbonization program.

While final experimental benchmarking is still underway, preliminary results suggest that the framework not only replicates known IL performance patterns but also uncovers previously untested candidates with favorable predicted properties. Importantly, the use of Pareto front analysis enables interpretable, multi-objective selection without relying on arbitrary weightings, surfacing ILs that strike optimal trade-offs between absorption capacity, viscosity, and regeneration energy. This approach supports flexible prioritization based on real-world process constraints and provides a rigorous, transparent basis for downstream experimental validation.

\textbf{Limitations.} Property predictions rely on data-driven models trained on incomplete and noisy experimental records. Although we opt for D-MPNN, baseline experiments reveal potential out-of-sample performance improvements with other predictive model options worth exploring and optimizing. Thermodynamic quantities are estimated assuming equilibrium behavior, which may not fully hold under dynamic process conditions. Moreover, viscosity predictions are only available at a single CO$_{2}$ loading state, and the effects of impurities, degradation, and phase transitions are not yet modeled.

Despite these constraints, the modularity of our framework offers a promising foundation for future work, like expanding the thermodynamic model to include pressure- and loading-dependence, incorporate toxicity and volatility constraints, and close the loop with experimental validation. 

%% file: sections/ack.tex
\begin{ack}
We gratefully acknowledge the support of SOCAR for their valuable operational insights, collaborative ideation, and strategic guidance throughout this project. Their contributions played a critical role in shaping the direction and relevance of this work to real-world carbon capture applications.
\end{ack}

%% file: sections/999_appendix.tex
\section{Data Curation \& Processing Details}
\label{sec:appendix-data}

\begin{itemize}
    \item Duplicate removal: Entries with identical (cation, anion, T, P) tuples but differing property values were compared. If relative deviation exceeded 5\%, the point was discarded.
    \item Extreme condition exclusion: To align with industrial applicability, we removed entries outside the relevant envelope: T > 420 K or P > 200 bar
    \item Physical consistency checks: (1) For solubility data, removed entries where solubility increased with increasing temperature at constant pressure, or decreased with increasing pressure at constant temperature—both violations of Henry’s law behavior. (2) For viscosity, discarded records showing increasing viscosity with temperature, which is physically implausible in ILs except under phase transitions.
\end{itemize}

\section{Directed Message Passing Neural Networks}
\label{sec:appendix-a}
The predictive models described in Section \ref{sec:dmpnn} are based on Directed Message Passing Neural Networks (D-MPNNs), which encode molecular graphs via information flow over directed bonds. The core design follows the formulation introduced by \citet{Yang2019} and implemented in the Chemprop framework \cite{Heid2023}.

\subsection{Molecular Graph Representation}
\label{sec:appendix-gnns}
Each IL ion (cation or anion) is treated as a graph G = (V, E), where:
\begin{itemize}
    \item $V$ is the set of atoms (nodes),
    \item $E \subseteq V \times V$ is the set of directed bonds (edges).
\end{itemize}
For each directed edge $e_{ij}$ from atom $i$ to atom $j$, we initialize a hidden feature vector $h_{ij}^{(0)}$ based on a featurization of the bond type, atom pair, and associated descriptors (e.g., aromaticity, hybridization).

\subsection{Message Passing Phase}
\label{sec:appendix-message-pass}
For T message passing steps, we update each directed edge feature using:
\begin{equation}
    h_{ij}^{(t+1)} = \text{ReLU}\left( W_m \cdot \sum_{k \in \mathcal{N}(i) \setminus j} h_{ki}^{(t)} + b_m \right)
\end{equation}
where, $\mathcal{N}(i) \setminus j$ is the set of neighbors of atom $i$ excluding $j$, $W_m$ and $b_m$ are learned weights and biases.

This formulation avoids loops and allows information to propagate along chemically meaningful bond directions.

\subsection{Readout Phase}
\label{sec:appendix-readout-phase}
After message passing, edge features are aggregated into atom-level representations and pooled to form a graph-level embedding:
\begin{equation}
    h_{\text{mol}} = \text{Aggregate}\left( \left\{ h_i \right\}_{i \in V} \right)
\end{equation}
where $h_i$ is a learned function of the incoming bond features to atom $i$, and the aggregate function is typically a sum or mean.

For ILs, we compute separate embeddings $h_{\text{cation}}$ and $h_{\text{anion}}$, and concatenate them to produce a single fixed-length vector for downstream regression:
\begin{equation}
    h_{\text{IL}} = [h_{\text{cation}} \,\|\, h_{\text{anion}}]
\end{equation}

This design retains the modularity of the IL representation while capturing substructural dependencies across both ions.

\section{Additional IL Candidate Analysis}
\label{sec:appendix-additional-analysis}

\begin{figure}[!h]
    \centering
    \includegraphics[width=0.8\linewidth]{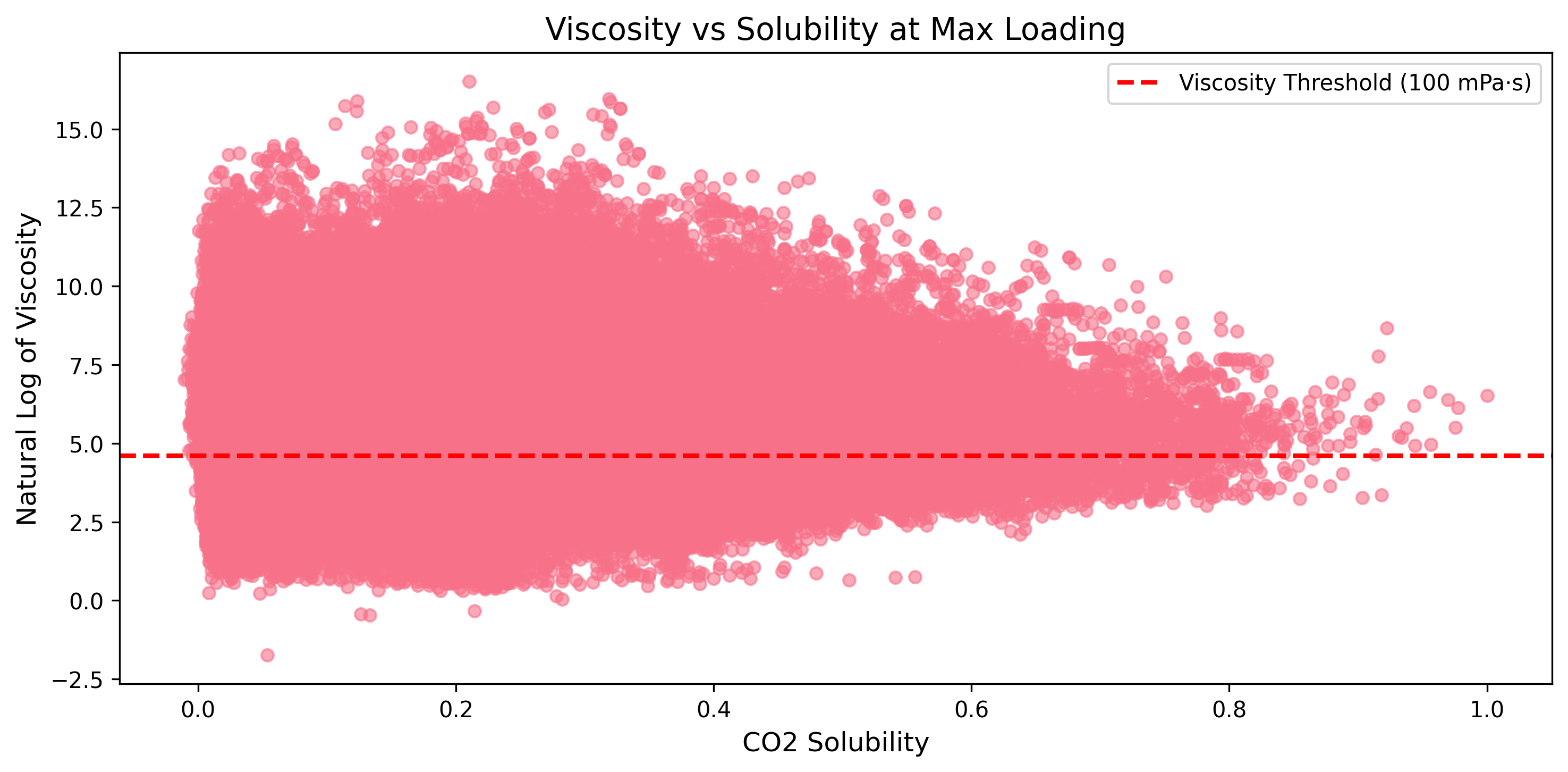}
    \caption{Predicted IL viscosity at various max CO$_2$ loading states. Plot reveals non-linear relationship between IL properties, with fewer promising ILs (high CO$_2$ solubility) predicted to operate at viscosities below 100 mPa·s, a generally accepted threshold for viable operations \cite{Gurkan2010}}
    \label{fig:loading_vs_viscosity}
\end{figure}

A scatter plot of maximum CO$_2$ loading against predicted viscosity (Figure~\ref{fig:loading_vs_viscosity}) reveals significant nonlinear coupling between these two properties. Many ILs with favorable max loading fail due to excessive viscosity at rich loading states, reaffirming the need for viscosity-aware solvent design.

\begin{figure}[!h]
    \centering
    \includegraphics[width=0.75\linewidth]{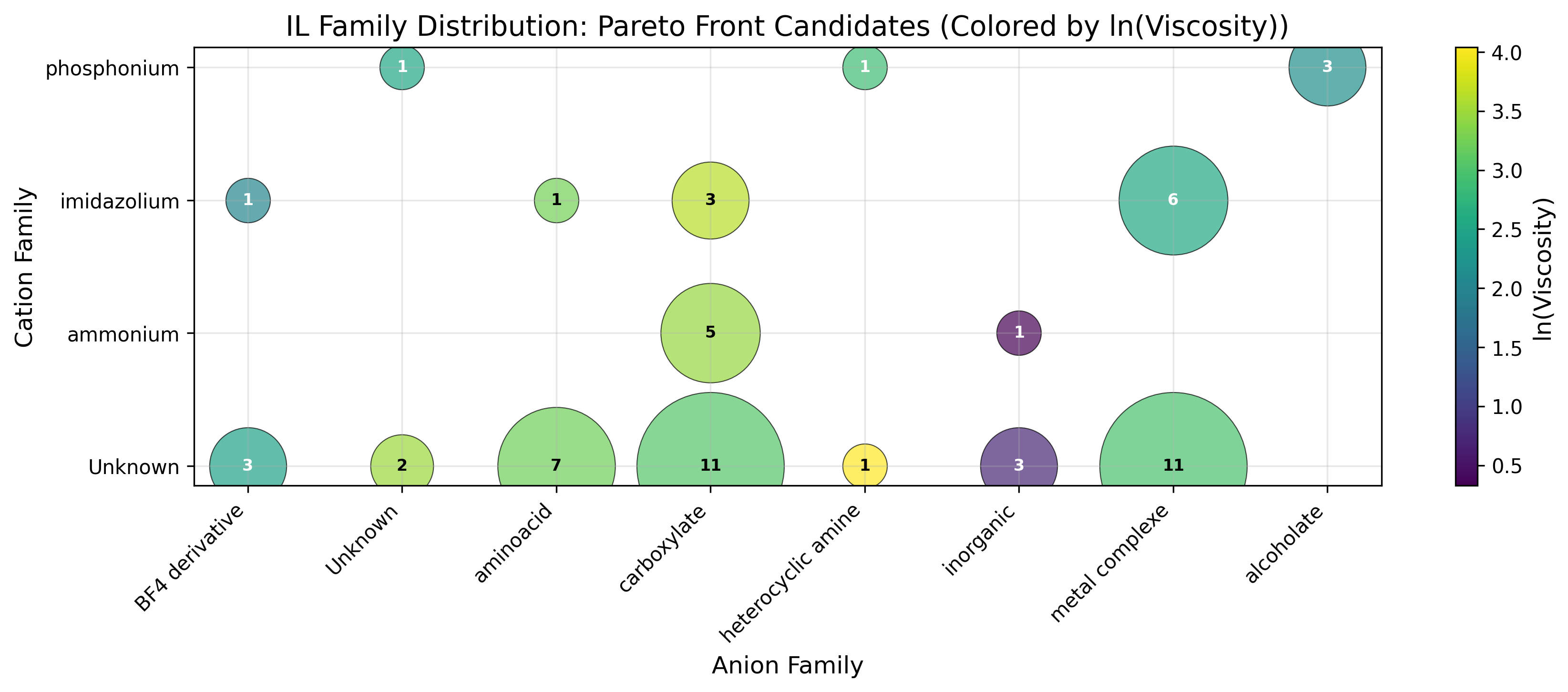}
    \caption{Bubble chart of Pareto front IL candidates grouped by cation and anion families. Bubble size is proportional to the number candidates. Color indicates average working capacity across the family combination candidates, where higher is better}
    \label{fig:pareto-family-performance-visc}
\end{figure}

\newpage

\section{Example Van't Hoff Fits}
\label{sec:appendix-vant-hoff-fits}

We observe a range in quality of Van't Hoff fits. Low quality fits suggest a strong temperature dependence for the enthalpy of absorption, possibly motivating the need for a polynomial fit to model a non-constant standard enthalpy.

\begin{figure}[!h]
    \centering
    \begin{minipage}[t]{0.50\linewidth}
        \centering
        \includegraphics[width=\linewidth]{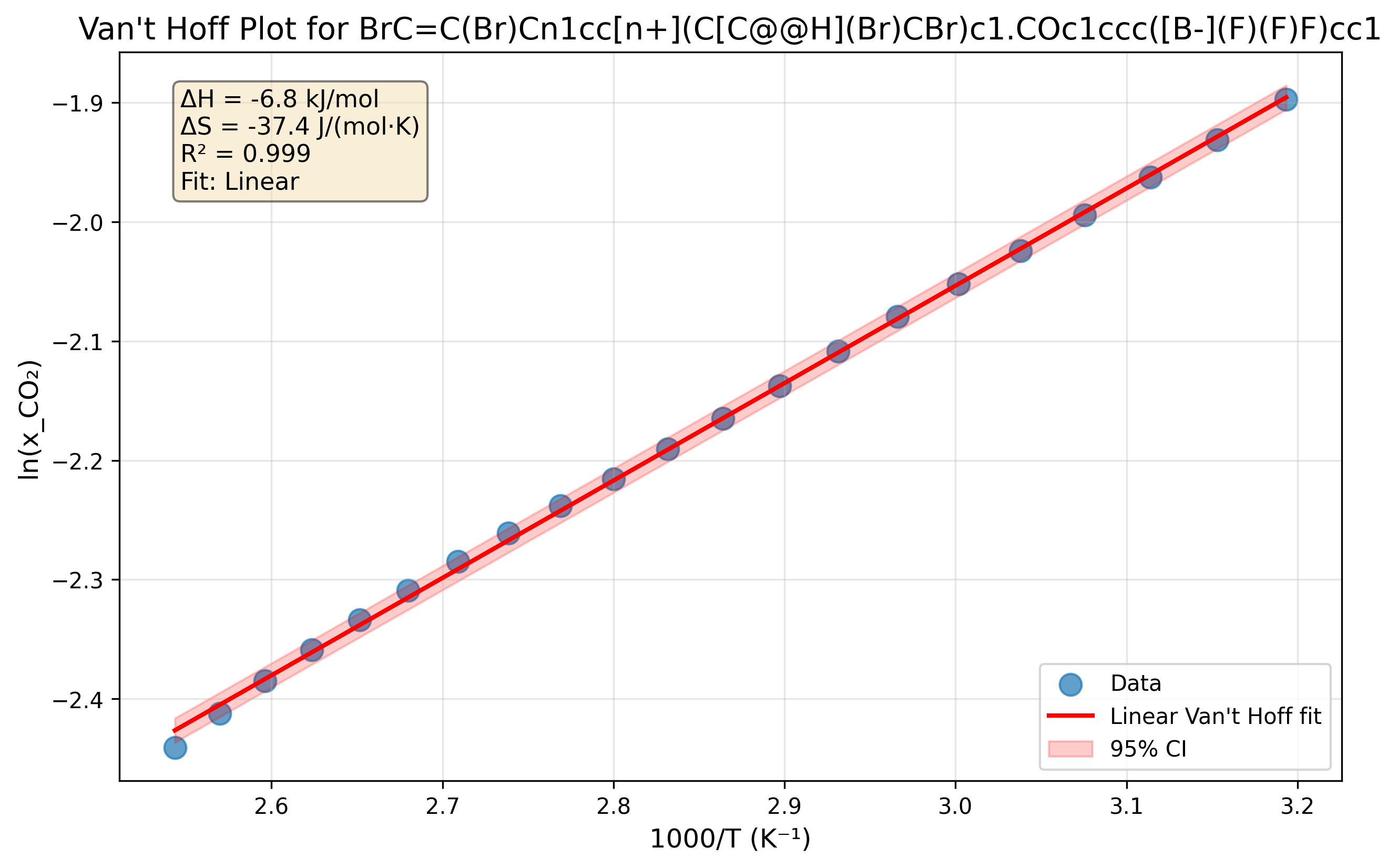}
    \end{minipage}
    \hfill
    \begin{minipage}[t]{0.48\linewidth}
        \centering
        \includegraphics[width=\linewidth]{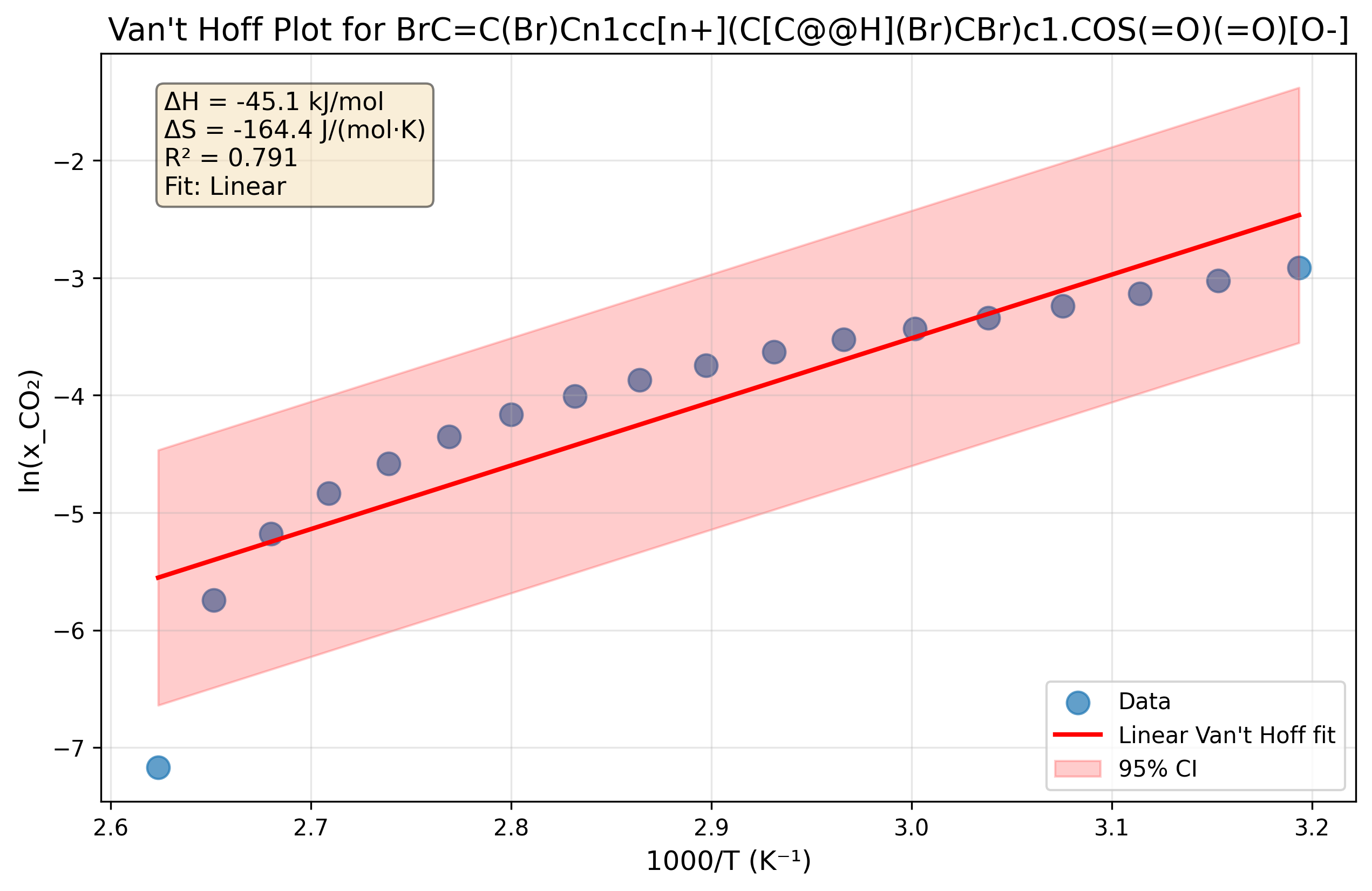}
    \label{fig:vant-hoff-2}
    \end{minipage}
\caption{Example Van't Hoff fits for ionic liquids. \textbf{Left.} Lower predicted heat of absorption with stronger fit. \textbf{Right.} Higher predicted heat of absorption with weaker fit.}
\label{fig:vant-hoff-plots}
\end{figure}